\documentclass[11pt]{article}
\usepackage{amsfonts}
\usepackage{mathrsfs}

\usepackage[latin1]{inputenc}
\usepackage{float}
\usepackage{amsmath,amssymb}
\usepackage{latexsym}
\usepackage{epstopdf}
\usepackage{geometry}  
\usepackage[active]{srcltx}
\usepackage{graphicx}
\usepackage{epstopdf}
\usepackage{booktabs}
\usepackage{multirow}
\usepackage{siunitx}

\usepackage[
bookmarks=true,         %generates bookmarks for all entries in the "Table of Contents"
bookmarksnumbered=true, %includes chapter-/header-/section-/subsection-/... -
colorlinks=true, pdfstartview=FitV, linkcolor=blue, citecolor=blue,
urlcolor=blue]{hyperref}

\newtheorem {theorem}{Theorem}[section]

\newtheorem{lemma}{Lemma}[section]

\newtheorem{remark}{Remark}[section]

\newcommand{\bi}[1]{\mbox{\boldmath{$ #1 $}}}

\def\E{{{\mathbb E}\,}}

\def\Var{{\mathop {{\rm Var\, }}}}

\begin{document}

%\begin{frontmatter}
%
\title{Jackknife Empirical Likelihood Methods for Gini Correlations and their Equality Testing}
%
%\author[mymainaddress]{Yongli Sang\corref{mycorrespondingauthor}}
%\cortext[mycorrespondingauthor]{Corresponding author}
%\ead{yongli.sang@louisiana.edu}
%
%\author[mysecondaryaddress]{Xin Dang}
%
%\author[mythirdaddress]{Yichuan Zhao}
%
%
%
%\address[mymainaddress]{Department of Mathematics, University of Louisiana at Lafayette, Lafayette, LA 70504, USA}
%\address[mysecondaryaddress]{Department of Mathematics, University of Mississippi, University, MS 38677, USA}
%\address[mythirdaddress]{Department of Mathematics and Statistics, Georgia State University, Atlanta, GA 30303, USA}

\author{Yongli Sang\textsuperscript{a}\thanks{CONTACT Yongli Sang. Email: yongli.sang@louisiana.edu}, Xin Dang\textsuperscript{b} and Yichuan Zhao\textsuperscript{c}}
\date{%
    \textsuperscript{a}Department of Mathematics, University of Louisiana at Lafayette, Lafayette, LA 70504, USA\\%
    \textsuperscript{b}Department of Mathematics, University of Mississippi, University, MS 38677, USA\\[2ex]%
     \textsuperscript{c}Department of Mathematics and Statistics, Georgia State University, Atlanta, GA 30303, USA\\[2ex]%
   % \today
}

\maketitle

\begin{abstract}
The Gini correlation  plays an important role in measuring dependence of random variables with heavy tailed distributions, whose properties are a mixture of Pearson's and Spearman's correlations. Due to the structure of this dependence measure, there are two Gini correlations between each pair of random variables, which are not equal in general. Both the Gini correlation and the equality of the two Gini correlations play important roles in Economics. In the literature, there are limited papers focusing on the inference of the Gini correlations and their equality testing.
In this paper, we develop the jackknife empirical likelihood (JEL) approach  for  the single Gini correlation, for testing the equality of the two Gini correlations, and for the Gini correlations' differences of two independent samples.  The standard limiting chi-square distributions of those jackknife empirical likelihood ratio statistics are established and used to construct confidence intervals, rejection regions, and to calculate $p$-values of the tests. 
Simulation studies show that our methods are competitive to existing methods in terms of coverage accuracy and shortness of confidence intervals, as well as in terms of power of the tests.  The proposed methods are illustrated in an application on a real data set from UCI Machine Learning Repository.

noindent  

\vskip.2cm 

\noindent {\bf Keywords:}
\noindent Jackknife empirical likelihood; Gini correlation; $U$-statistics; Wilks' theorem; Test\\
%\MSC[2010] 62G35\sep  62G20

\end{abstract}

\section{Introduction}

The Gini correlation has been  used  in a wide range of fields since proposed in 1987 (\cite{Schechtman87}).  In the field of economic data analysis, the Gini correlation enables us to test whether an asset increases or decreases the risk of the portfolio (\cite{Schechtman99}), and can be used to build the relationship between the family income and components of income (\cite{Schechtman87}); in plant systems biology, the Gini correlation  can compensate for the shortcomings of  popular correlations in inferring regulatory relationships in transcriptome analysis (\cite{Ma2012}); it has also been widely used in all branches of modern signal processing (\cite{Xu10}). 

Let $X$ and $Y$  be two non-degenerate random variables  with continuous marginal distribution functions $F$ and $G$, respectively, and a joint distribution function $H$. Then two Gini correlations are defined as 
\begin{align}\label{eqn:rho}
&\gamma(X,Y):=\frac{\mbox{cov}(X, G(Y))}{\mbox{cov}(X,F(X))} \;\;\;\text{and }\;
\gamma(Y, X):=\frac{\mbox{cov}(Y, F(X))}{\mbox{cov}(Y,G(Y))}
\end{align}
to reflect different roles of $X$ and $Y.$ 
The representation of Gini correlation $\gamma(X, Y )$ indicates that it has mixed properties of those of the Pearson and Spearman correlations, and  thus complements these two correlations (\cite{Schechtman87}, \cite{Schechtman99}, \cite{Schechtman03}). The two Gini correlations in (\ref{eqn:rho}) are not symmetric in X and Y in general. The equality of the two Gini correlations can be involved in many procedures in Economics. For example,  it can be applied to determine the similarity in two popular methodologies for constructing portfolios, the MV and MG (\cite{Schechtman07}), and the equality of the two Gini correlation between the return on each asset and the return on the portfolio is the necessary condition of the statement that all the Security Characteristic curves are linear (\cite{Schechtman07}), that is, a rejection of the hypothesis on the equality of Gini correlations is a rejection of the assumption that all the Security Characteristics curves are linear. Therefore, to understand the Gini correlation and to test the equality of the two Gini correlations are essential. In the paper, we develop a procedure to estimate the Gini correlation and to test the equality of the two Gini correlations. To the best of our knowledge, there is no nonparametric approaches  to infer the Gini correlations. 

As a nonparametric method, the empirical likelihood (EL) method introduced by Owen (\cite{Owen1988}, \cite{Owen1990}) has been used heuristically  for constructing confidence intervals.  It combines the effectiveness of likelihood and the reliability of nonparametric approach. On the computational side,  it involves a maximization of  the nonparametric likelihood supported on data subject to some constraints. If these constraints are linear, the computation of the EL method is particularly easy. However,  EL loses this efficiency when some nonlinear constraints are involved. To overcome this computational difficulty,  Wood $et$ $al.$ (\cite{Wood1996}) proposed a sequential linearization method by linearizing the nonlinear constraints. However, they did not provide the Wilks' theorem and stated that it was not easy to establish.  Jing {\em et al.} (\cite{Jing2009}) proposed the jackknife empirical likelihood (JEL) approach. The JEL method transforms  the maximization problem of the EL with nonlinear constraints to the simple case of EL on the mean of  jackknife pseudo-values, which is very effective in handling one and two-sample $U$-statistics.  Wilks' theorems for one and two-sample $U$-statistics are established.  This approach has attracted statisticians' strong interest in a wide range of fields due to its efficiency, and many papers are devoted to the investigation of the method, for example, \cite{Liu2015}, \cite{Peng2012}, \cite{Feng2012}, \cite{Zhao2016}, \cite{Wang2013}, \cite{Li2016}, \cite{Li2011} and so on. However, theorems derived in \cite{Jing2009} are limited to a simple case of the $U$-statistic but the Gini correlation  cannot be estimated by a $U$-statistic, which does not allow us to  apply the results of \cite{Jing2009} directly. 
However,  it can be estimated by a functional of multiple $U$-statistics (\cite{Schechtman87}). Due to this specific form of the Gini correlation, we propose a novel $U$-statistic type functional and a JEL-based procedure with the $U$-structured estimating function is applied for the Gini correlation. 
And this approach may work for making an inference about some difference functions of multiple $U$-statistic structure with nuisance parameters involved.

In the test
\begin{align}\label{hypotest}
H_0: \Delta=0 \;\;  \text{vs}  \;\; H_a: \Delta \neq 0,
\end{align}
where $\Delta =\gamma(X,Y)-\gamma(Y,X)$,
the natural empirical estimator $\hat{\Delta}$ of $\Delta$
%$\hat{\Delta}=\hat{\gamma}(X,Y)-\hat{\gamma}(Y,X)$ of $\Delta$
 is a function of 4 dependent $U$-statistics. 
Based on $U$-statistics theorem,  $\hat{\Delta}$, will, after appropriate normalization, have a limiting normal distribution. However, the asymptotic variance is complicated to calculate.  
%Schechtman $et$ $al.$ (\cite{Schechtman07}) recommended  an estimation of the variance based on the jackknife method (\cite{Shao96}). 
In the present paper, by proposing a new $U$-statistic type functional system, we avoid estimating the asymptotic variance to do the test.  However,  only a part of parameters are being interested. When only a part of parameters are of interest, Qin and Lawless (\cite{Qin1994}) proposed to use a profile empirical likelihood method which  is also an important tool to transform nonlinear constraints to some linear constraints by introducing  link nuisance parameters. However, the profile EL could be computationally costly. Hjort, McKeague and Van Keilegom (\cite{Hjort09}) proposed to reduce the computational complexity by allowing for plug-in estimates of nuisance parameters in estimating equations with the cost that the standard Wilks' theorem may not hold. Li {\em et al.} (\cite{Li2011}) proposed a jackknife plug-in estimate in terms of a function of interested parameters so that their EL still have standard chi-square distributions. However, we cannot take advantage of their method since the parameters of interest in this paper are estimated by solving estimating functions with $U$-statistics structure. We cannot apply theoretical results of the profile JEL method in \cite{Li2016}, either. Li, Xu and Zhao (\cite{Li2016}) developed a JEL-based inferential procedure for general $U$-structured estimating functions. It requires the condition that kernel functions are bounded both in the sample space and in the parameter space.  Under merely second order moment  assumptions, we establish the Wilks' theorem for the jackknife empirical log-likelihood ratio for $\Delta$. The computation is also easy since a simple plug-in estimate of the nuisance parameter is used.

It is often of considerable interest to compare the Gini correlations from two independent populations. For instance, Lohweg $et$ $al.$ (\cite{Lohweg2013}) constructed adaptive wavelets for the analysis of different print patterns on a banknote and made it possible to use mobile devices for banknote authentication. After the wavelet transformations,  there are four continuous variables: variance, skewness, kurtosis and entropy of wavelet transformed images. It is natural to ask what are correlations of each pair of the above variables. Are there any differences between the Genuine banknotes and Forgery banknotes? One of the main goals of this paper is to develop the JEL method for comparing the Gini correlations for independent data sets.

The remainder of the paper is organized as follows.  In Section 2,  we develop the JEL method for the Gini correlations.  The JEL method for testing the equality of Gini correlations  is proposed in Section 3.  In Section 4, we consider the JEL method for comparing Gini correlations for two samples.   Following the introduction of methods in each section, simulation studies are conducted to compare our JEL methods with some existing methods.  A real data analysis is illustrated in Section 5. Section 6 concludes the paper with a brief summary. All proofs are reserved to the Appendix.

\section{JEL for the Gini correlation}
The Gini correlation
%is based on the covariance between two random variables where one variable is taken in its variate values while the other is ranked.  It 
has a mixed property of the Pearson correlation and the Spearman correlation: (1) If $X$ and $Y$ are statistically independent then $\gamma(X, Y)=\gamma(Y, X)=0$; (2)  $\gamma(X, Y)$ is invariant under all strictly increasing transformations of $Y$ or under changes of scale and location in $X$; (3)  $-1 \leq \gamma(X, Y) \leq 1$ and (4) if $Y$ is a monotonic increasing (decreasing) function of $X$, then both $\gamma(X, Y)$ and $\gamma(Y, X)$ equal +1 (-1). From \cite{Schechtman87}, $\gamma(X, Y)$ in (\ref{eqn:rho}) can be written in the form as below 
\begin{align}\label{eqn:rho12}
\gamma(X,Y)=\frac{\mathbb{E}h_{1}((X_1, Y_1),(X_2, Y_2))}{\mathbb{E}h_{2}((X_1, Y_1),(X_2, Y_2))},
\end{align}
where $(X_1, Y_1)^T$ and $(X_2, Y_2)^T$ are independent copies of $(X, Y)^T$,  
\begin{align}\label{h1}
h_{1}((x_1, y_1), (x_2, y_2))= \frac{1}{4}[(x_1-x_2)I (y_1>y_2) +(x_2-x_1)I(y_2>y_1)]
\end{align} and \begin{align}\label{h2} h_{2}((x_1, y_1), (x_2, y_2))=\frac{1}{4}|x_1-x_2|.\end{align} 
Then given an i.i.d. data set  $\mathcal{Z}=\{\bi Z_1, \bi Z_2, ..., \bi Z_n\}$, $n \geq 2$ with $\bi Z_i=(X_i, Y_i)^T, i=1, ..., n$, the Gini correlation $\gamma(X, Y)$ can be estimated by a ratio of two $U$-statistics 
\begin{equation}\label{eqn:rhohat}
\hat{\gamma}(X, Y)=\frac{U_1}{U_2} =  \frac{ {n \choose 2}^{-1}\sum_{1\le i<j \le n} \frac{1}{4}[(X_i-X_j)I (Y_i>Y_j) +(X_j-X_i)I(Y_j>Y_i)]}{ {n \choose 2}^{-1}\sum_{1\le i<j\le n}\frac{1}{4}|X_i-X_j|}
\end{equation}
 with the kernel of $U_1$ being $h_1$ and the kernel  of $U_2$ being $h_2$. 
 \begin{remark}
A direct computation of $U$-statistics is time-intensive with complexity $O(n^2)$. Rewriting $U_1$ and $U_2$ as linear combinations of order statistics reduces the computation to $O(n\log n)$. That is, $U_1=\frac{1}{4 {n \choose 2}} \sum_{i=1}^n (2i-1-n)X_{(Y_{(i)})}$ and $U_2=\frac{1}{4 {n \choose 2}} \sum_{i=1}^n (2i-1-n)X_{(i)}$, where $X_{(i)}$ is the $i^{th}$ order statistics of $X_1, X_2, ..., X_n$ and $X_{(Y_{(i)})}$ is the $X$ that belongs to $Y_{(i)}$ (\cite{Schechtman87}). 
 \end{remark}
 By $U$-statistic theory, the asymptotic normality of the estimator (\ref{eqn:rhohat}) for $\gamma(X ,Y)$ (\cite{Schechtman87}, \cite{Sang2016}) is:
\begin{align}\label{3.5}
\sqrt{n}(\hat{\gamma}(X, Y)-\gamma(X, Y))\stackrel{d} \rightarrow N(0, v_{\gamma}) \;\;\;as \;\;\; n\rightarrow \infty,
\end{align}
with  
\begin{align}\label{asyv}
v_{\gamma}=({4}/{\theta^2_2}) \zeta_1(\theta_1)+({4\theta^2_{1}}/{\theta^4_2})\zeta_2(\theta_2)-({8\theta_{1}}/{\theta^3_2}) \zeta_3(\theta_1,\theta_2), 
\end{align}
where $$\theta_1=\mbox{cov}(X, G(Y)), \;\;\theta_2=\mbox{cov}(X, F(X)),$$
$$\zeta_1(\theta_1)=\E_{\bf {z}_1}\left\{ \E_{\bf{ z}_2} [h_1(\bf{Z}_1,\bf{Z}_2)]\right\}^2-\theta_1^2,$$ 
$$\zeta_2(\theta_2)=\E_{\bf {z}_1}\left \{\E_{\bf{ z}_2}[h_2(\bf{ Z}_1,\bf{ Z}_2)]\right\}^2-\theta_2^2 $$ and 
$$\zeta_3(\theta_1, \theta_2)=\E_{\bf{ z}_1}\left \{\E_{\bf{ z}_2}[h_1(\bf{ Z}_1,\bf{ Z}_2)]  \E_{\bf{ z}_2}[h_2(\bf{ Z}_1,\bf{ Z}_2)]\right\}-\theta_1\theta_2.$$  In particular, under a bivariate normal distribution with correlation $\rho$, Xu $et$ $al.$ (\cite{Xu10}) provided an explicit formula of $v_{\gamma}$, given by $v_{\gamma} = \pi/3 +(\pi/3+4\sqrt{3})\rho^2-4\rho \arcsin (\rho/2)-4\rho^2\sqrt{4-\rho^2}$. 
 However, the asymptotic variance $v_{\gamma}$ is difficult to obtain for the non-normal distributions and an estimate of  $v_{\gamma}$ is needed either by a Monte Carlo simulation or based on the jackknife method. 
Let $\hat {\gamma}_{(-i)}$ be the jackknife pseudo value of the Gini correlation estimator $\hat {\gamma}(X,Y)$ based on the sample with the $i^{th}$ observation deleted. Then the jackknife estimator of (\ref{asyv}) is
\begin{align}\label{jelrho}
\hat{v}_{\gamma} =\frac{n-1}{n}\sum_{i=1}^n ({\hat{\gamma}}_{(-i)} -\bar{\hat{\gamma}}_{(\cdot)})^2
\end{align}
where $\bar{\hat{\gamma}}_{(\cdot)} =1/n \sum_{i=1}^n {\hat{\gamma}}_{(-i)},$ see \cite{Shao96}.

In this section, we utilize the jackknife approach combining with the EL method to make inference on the Gini correlation.

\subsection{JEL for the Gini correlation}
%\bigskip\textbf{}
Without loss of generality, we consider the case for $\gamma(X, Y)$, and  the procedure for $\gamma(Y, X)$ will be similar. For simplicity, we use  $\gamma$ to  denote $\gamma(X, Y)$ in this section. 
%Jing $et$ $al.$(\cite{Jing2009}) proposed the JEL approach and founded the Wilks' theorems for $U$-statistics. However, the estimator in (\ref{eqn:rhohat}) of the Gini correlation $\gamma$ is not a $U$-statistic, which do not allow us to  apply the results of \cite{Jing2009} directly. 
Define a novel $U$-statistic type functional as 
\begin{align} \label{eqn: U_ngamma}
U_{n}(\gamma)={n \choose 2}^{-1}\sum_{1\le i<j \le n}h((X_{i}, Y_{i}), (X_{j}, Y_{j});  \gamma),
\end{align}
where 
\begin{align}\label{eqn:kernel}
h((x_1,y_1), (x_2,y_2); \gamma)=h_2((x_1,y_1), (x_2,y_2)) \cdot \gamma-h_1((x_1,y_1), (x_2,y_2))
\end{align} 
with $h_1(\cdot)$ and $h_2(\cdot)$ being given by (\ref{h1}) and (\ref{h2}), respectively. 
By (\ref{eqn:rho12}), we have $\E h((X_{i},Y_{i}), (X_{j},Y_{j});\gamma)=0$. 
To apply the JEL to $U_n(\gamma)$, we  define the jackknife pseudo sample as 
\begin{align*}
\hat{V}_{i}(\gamma)=n U_{n}(\gamma)-(n-1)U^{(-i)}_{n-1}(\gamma),
\end{align*}
where $U^{(-i)}_{n-1}(\gamma)$ is based on the sample  with the $i^{th}$ observation $(X_{i},Y_{i})^T$ being deleted. It can be easily shown that  $\mathbb{E}[\hat{V}_{i}(\gamma)]=0$ and
\begin{align*}
U_{n}(\gamma)=\frac{1}{n}\sum_{i=1}^n\hat{V}_{i}(\gamma).
\end{align*}
Let $\bold{p}=(p_1,...,p_n)$ be nonnegative numbers such that $\sum_{i=1}^{n}p_i=1.$  Then following the standard empirical likelihood method for a univariate mean over the jackknife pseudo-values (\cite{Owen1988}, \cite{Owen1990}), we define the JEL ratio at $\gamma$ as
\begin{align*} 
R(\gamma)=\max \left \{\prod_{i=1}^{n}(np_i): p_i \ge 0, i=1,...,n;  \sum_{i=1}^n p_i=1; \sum_{i=1}^n p_i \hat{V}_{i}(\gamma)=0 \right \}.
\end{align*}
Utilizing the standard Lagrange multiplier technique, the jackknife empirical log-likelihood ratio at $\gamma$ is
\begin{align*}
\log R(\gamma)=-\sum_{i=1}^{n}\log[1+\lambda \hat{V}_{i}(\gamma) ],
\end{align*}
where $\lambda=\lambda(\gamma)$ satisfies 
\begin{align*}
\frac{1}{n}\sum_{i=1}^n \frac{ \hat{V}_{i}(\gamma)}{1+\lambda \hat{V}_{i}(\gamma)}=0.
\end{align*}

 Define 
$g((x, y);\gamma)=\E h((x, y), (X_2, Y_2);\gamma)$ and $\sigma^{2}_{g}(\gamma)=\Var (g( (X_1, Y_1); \gamma))$. Then we have the following Wilks' theorem with only the assumption of the existence of second moments:
\begin{theorem}\label{wilkrho}
Assume $\E X^2_1<\infty,$ ,  $\E Y^2_1<\infty$ and $\sigma^{2}_{g}(\gamma)>0$. Then we have 
$$ -2 \log R(\gamma)\stackrel{d}{\rightarrow} \chi^{2}_1, \;\;\;\text{as $n \to \infty$}.$$
\end{theorem}
Based on the theorem above, a  $100(1-\alpha)\%$ jackknife empirical likelihood confidence interval for $\gamma$  can be constructed  as
\begin{align*}
I_{\alpha} = \{ \tilde{\gamma}: -2 \log \hat{R}(\tilde{\gamma})\le \chi^2_{1, 1-\alpha} \},
\end{align*}
where $\chi^2_{1, 1-\alpha}$ denotes the $100(1-\alpha)\%$ quantile of the chi-square distribution with one degree of freedom, and $ \log \hat{R}(\tilde{\gamma})$ is the observed empirical log-likelihood ratio at $\tilde \gamma$.

%\begin{remark}
%Li $et$ $al.$ (\cite{Li2016}) established the Wilks' theorem for a general U-type empirical likelihood ratio under a strong condition that the kernel $h$'s are uniformly bounded in both variables and parameters. Here we only assume the existence of the second moments.
%\end{remark}

%\begin{remark}
%For the Gini correlation between $Y$ and $X$, $\gamma(Y,X)$, we have similar results. 
%\end{remark}

In application, an under-coverage problem may appear when the sample size is relatively small. In order to improve coverage probabilities, we utilize the adjusted empirical likelihood method (\cite{Chen2008}) by adding one more  pseudo-value 
$$\hat{V}_{n+1}(\gamma)=-\frac{a_n}{n}\sum^{n}_{i=1}\hat{V}_{i}(\gamma),$$ 
where $a_n=o_{p}(n^{2/3}).$ Under the recommendation of \cite{Chen2008}, we take  $a_n=\max (1, \log (n)/2).$

\subsection{Empirical performance}
%\bigskip\textbf{}
To evaluate the empirical  performance of our JEL methods (denoted as `JEL$\gamma_1$', `JEL$\gamma_2$' for $\gamma(X, Y)$, $\gamma(Y, X)$, respectively), we conduct a simulation study.  
Another purpose is to examine whether the adjusted JEL methods (denoted as `AJEL$\gamma_1$', `AJEL$\gamma_2$' ) can make an improvement over the JEL method for small sample sizes. The interval estimators for the Gini correlations based on the asymptotical normality of (\ref{3.5}) with variance calculated by (\ref{asyv}) are denoted as `$\gamma_{1}$AV' and `$\gamma_{2}$AV', while `$\gamma_{1}$J'  and `$\gamma_{2}$J' to denote the methods using (\ref{jelrho}) to estimate $v_{\gamma}$.  Similar notions for the different method in the following sections will be used. 

We also present the results for the Pearson's correlation and denote it as `$\rho_p$'. The limiting distribution of the regular sample Pearson correlation coefficient $\hat{\rho}_p$ is normal:
 $\sqrt{n}(\hat{\rho}_p-\rho) \stackrel{d}{\rightarrow} N(0, v_p)\;\; as\;\; n \to \infty$, 
%\begin{align*}
%\sqrt{n}(\hat{\rho}_p-\rho) \stackrel{d}{\rightarrow} N(0, v_p)\;\; as\;\; n \to \infty,  
%\end{align*}
where
\begin{align}\label{vp}
v_p=(1+\frac{\rho^2}{2})\frac{\sigma_{22}}{\sigma_{20}\sigma_{02}}+\frac{\rho^2}{4}(\frac{\sigma_{40}}{\sigma^2_{20}}+\frac{\sigma_{04}}{\sigma^2_{02}}-\frac{4 \sigma_{31}}{\sigma_{11}\sigma_{20}}-\frac{4 \sigma_{13}}{\sigma_{11}\sigma_{02}}), %\label{pearson}
\end{align}
and $\sigma_{kl}=\E[(X-\E X)^{k}(Y-\E Y)^l]$, see for example \cite{Witting95}. The Pearson correlation estimator requires a finite fourth moment on the distribution to evaluate its asymptotic variance. For bivariate normal distributions, the asymptotic variance $v_p$ simplifies to $(1-\rho^2)^2$. For other distributions rather than the normal, the asymptotic variance may be estimated by a Monte Carlo simulation or by a jackknife variance method. We do not include another popular correlation Kendall $\tau$  in the simulation. Its performance is referred   to \cite{Sang2016}.

We generate 3000 samples of two different sample sizes ($n=20, 200$) from two different bivariate distributions, namely, normal and $t(5)$,  with the scatter matrix $\bf\Sigma=\begin{pmatrix}1 \;\;\;&2\rho\\ 2\rho\;\;\;&4\end{pmatrix}$. Without loss of generality, we consider only cases of $\rho>0$ with $\rho=0.1, 0.5, 0.9$.  For each simulated data set, $90\%$ and $95\%$ confidence intervals  are calculated using different methods. We repeat this procedure 30 times. The average coverage probabilities and average lengths of confidence intervals as well as their standard deviations (in parenthesis) are presented in Tables \ref{tab1:covprho}, \ref{tab2:covprho}.

\begin{table}[!htbp]
%\center
%\begin{sidewaystable}[thb]
\center
%\tiny
\scriptsize
%\footnotesize
\caption{Coverage probabilities (standard deviations) and average lengths (standard deviations) of the Gini correlations' interval estimators from a variety of methods under bivariate normal distributions. }
\label{tab1:covprho}

\begin{tabular}{ll|c   c|c  c }
\hline \hline
 && \multicolumn{2}{c|}{$n=20$}&
 \multicolumn{2}{c}{$n=200$}\\ 
{$\rho$}&{Method}&{$1-\alpha=0.90$}   & {$1-\alpha=0.95$}& {$1-\alpha=0.90$}   & {$1-\alpha=0.95$}   \\
&&CovProb  Length &CovProb  Length&CovProb  Length&CovProb  Length\\
\hline

&JEL$\gamma_{1}$ &.880(.005) .642(.002) &.931(.005) .799(.004)   
  & .900(.005) .208(.002) &.950(.003) .230(.000) \\
 
    &  AJEL$\gamma_{1}$  &.907(.005)  .710(.003) &.951(.004)  .937(.005)    
     & .905(.005) .210(.000) &.953(.003)  .232(.000) \\
 
 &$\gamma_{1}$J &.869(.006)  .812(.002)& .916(.006)  .968(.004)    
  & .898(.005) .238(.000)& .947(.004) .284(.000) \\
 
& $\gamma_{1}$AV&.886(.005)  .748(.000) &.945(.005)  .891(.000)     
& .900(.005) .237(.000) &.951(.004) .282(.000) \\
  
$\rho=0.1$  & JEL$\gamma_{2}$  &.880(.005)  .642(.002) &.931(.006)  .798(.006)  
   & .900(.005) .208(.000)& .950(.003) .230(.000) \\
   
& AJEL$\gamma_{2}$  &.908(.003)  .709(.003) &.951(.005)  .936(.007)
     & .904(.005) .209(.000) &.953(.003) .232(.000) \\
     
 &$\gamma_{2}$J&.869(.004)  .811(.003) &.916(.006)  .968(.005)     
 & .897(.005) .238(.000) &.947(.003) .283(.000) \\
 
& $\gamma_{2}$AV&.886(.004)  .748(.000) & .945(.005)  .891(.000)  
   & .900(.005) .237(.000) &.951(.003) .282(.000) \\
   
 &$\rho_{p}$  &.889(.004)  .728(.000) &.947(.006)  .868(.000) 
     & .899(.005) .230(.000) &.950(.004) .274(.000) \\
 \hline

       &JEL$\gamma_{1}$ &.876(.007)  .506(.002)& .925(.006)  .572(.004)     
       & .900(.006)  .156(.000) &.950(.004)  .171(.000) \\
 
      &AJEL$\gamma_{1}$  &.903(.006)  .530(.003)& .944(.005)  .673(.005)
           & .905(.006) .157(.000) &.953(.004)  .173(.000) \\
 
 &$\gamma_{1}$J &.867(.007)  .634(.003) &.908(.007)  .756(.005)     
 & .898(.006)  .182(.000) &.947(.004) .217(.000) \\
 
& $\gamma_{1}$AV&.886(.005)  .568(.000) &.937(.005)  .677(.000) 
    & .898(.006)  .180(.000) &.949(.004) .214(.000) \\
  
 $\rho=0.5$  &JEL$\gamma_{2}$  &.876(.006)  .506(.002) &.926(.006)  .572(.004)     
 & .900(.006)  .156(.000) &.950(.003)  .171(.000) \\
 
 &AJEL$\gamma_{2}$  &.903(.006)  .530(.002) &.945(.005)  .673(.004)    
  & .905(.006)  .157(.000) &.953(.003)  .173(.000) \\
 
 &$\gamma_{2}$J&.865(.006)  .634(.003)& .908(.007)   .757(.005)
      & .897(.006)   .182(.000)& .946(.004)   .217(.000) \\
      
& $\gamma_{2}$AV&.886(.005)   .568(.000)& .937(.005)  .677(.000)   
  & .899(.006)  .180(.000) &.949(.003)   .214(.000) \\
  
 &$\rho_{p}$  &.893(.005)  .552(.000)& .941(.005)  .657(.000)     
 & .900(.005)  .175(.000) &.950(.003)  .208(.000) \\

\hline

      & JEL$\gamma_{1}$ &.874(.005)  .144(.001)    & .919(.005)  .183(.002)             
      & .899(.005)  .040(.000)& .948(.005) .044(.000) \\
 
 &AJEL$\gamma_{1}$  &.898(.005)  .152(.002)           & .936(.005)  .254(.003)             
 & .903(.006)  .040(.000) & .951(.005)   .044(.000) \\

 &$\gamma_{1}$J &.857(.007)  .185(.002)           &     .892(.005) .220(.002)    
  & .897(.006) .048(.000) &.942(.005) .057(.000) \\

& $\gamma_{1}$AV&.876(.006)  .146(.000) &     .919(.006)  .174(.000)     
& .894(.006)  .046(.000) & .943(.005)  .055(.000) \\

 $\rho=0.9$  &JEL$\gamma_{2}$  &.874(.007)  .143(.001)        &        .919(.004)  .184(.002)
             & .899(.006)  .040(.000) &.949(.005)  .044(.000) \\
             
& AJEL$\gamma_{2}$  &.898(.006)  .151(.002)  &  .937(.004)  .254(.003)
     & .903(.006)  .040(.000)  &.952(.005) .044(.000) \\
     
  &$\gamma_{2}$J&.858(.007)  .184(.002)       &       .892(.006)  .220(.002)
       & .896(.006)  .048(.000)& .943(.006) .057(.000) \\
       
&  $\gamma_{2}$AV&.876(.007)  .146(.000)   &          .918(.006)  .174(.000)  
   & .894(.006)   .046(.000) &.944(.005)  .055(.000) \\

& $\rho_{p}$  &.894(.006)  .140(.000)        &        .929(.005)   .167(.000)  
   & .900(.006)   .044(.000)  &.947(.004)  .053(.000) \\

\hline\hline
\end{tabular}
\end{table}

  Under elliptical distributions including normal and $t$ distributions, the two Gini correlations and the Pearson correlation are equal to the linear correlation parameter $\rho$, that is, $\gamma(X, Y)=\gamma(Y, X)=\rho_P=\rho$ (\cite{Sang2016}). Thus, all the listed methods in Table \ref{tab1:covprho} and Table \ref{tab2:covprho} are for the same quantity $\rho$. 
    
From Table \ref{tab1:covprho}, we observe that under the bivariate normal distribution, all methods keep good coverage probabilities when sample size is large ($n=200$) but the JEL methods produce the shortest intervals for all $\rho$ values.  It even behaves better than the $\rho_P$ method which is asymptotically optimal under normal distributions. $\gamma$AV ($\gamma_1$AV, $\gamma_2$AV) methods are slightly better than $\gamma$J ($\gamma_{1}$J, $\gamma_{2}$J) methods but not good as $\rho_P$ method. Note that the lengths of $\gamma_1$AV and $\gamma_2$AV methods are same, also standard deviations of confidence interval lengths for $\gamma$AV, $\gamma$J and $\rho_P$ methods are always 0.
 When the sample size is relatively small ($n=20$), our JEL method always produces better coverage probabilities and shorter confidence intervals compared with $\gamma$J method, and performs better than $\gamma$AV method when $\rho$ is relatively large ($\rho=0.9$). All the JEL, $\gamma$J and $\gamma$AV methods present slight under-coverage problems. However, the adjusted JEL method improves the under-coverage problems effectively and keeps shorter intervals.

\begin{table}[!htbp]
%\center
%\begin{sidewaystable}[H]
\center
\tiny
\scriptsize
%\footnotesize
\caption{Coverage probabilities (standard deviations) and average lengths (standard deviations) of the Gini correlations' interval estimators from a variety of methods  under bivariate $t(5)$ distributions.}
\label{tab2:covprho}
\begin{tabular}{ll|c   c|c  c }
\hline \hline
 && \multicolumn{2}{c|}{$n=20$}&
 \multicolumn{2}{c}{$n=200$}\\
{$\rho$}&{Method}&{$1-\alpha=0.90$}   & {$1-\alpha=0.95$}& {$1-\alpha=0.90$}   & {$1-\alpha=0.95$}   \\
&&CovProb  Length &CovProb  Length&CovProb  Length&CovProb  Length\\
\hline

       &JEL$\gamma_{1}$ &.853(.007)  .467(.003) 
       &.910(.005) .538(.002)     
       & .892(.004)  .210(.001) 
       &.944(.003) .239(.001) \\
 
     & AJEL$\gamma_{1}$  &.887(.006)  .505(.003)
      &.936(.005)  .585(.002)     
      & .897(.004)  .212(.001) 
      &.948(.003)  .242(.001) \\
      
  &$\gamma_{1}$J &.853(.007)   .912(.005)
   &.901(.006)  1.08(.005)     
   & .893(.004)  .285(.000) 
   &.944(.004)  .340(.001) \\
 
 &$\gamma_{1}$AV&.901(.006)  .894(.000) 
 &.957(.003)  1.07(.000)     
 & .897(.004) .283(.000) 
 &.948(.003)  .337(.000) \\
  
 $\rho=0.1$&JEL$\gamma_{2}$  &.854(.007)  .468(.003)
  &.911(.004)  .538(.002)     
  & .892(.005) .210(.001) 
  &.944(.003)  .239(.001) \\
  
 &AJEL$\gamma_{2}$  &.887(.006)  .505(.003)
  & .937(.004)  .584(.002)     
  & .897(.005)  .213(.001) 
  &.947(.003) .242(.001) \\
  
&$\gamma_{2}$J&.853(.006)  .909(.005) 
&.901(.005)  1.08(.005)     
& .893(.005)  .285(.001) 
&.943(.003)  .340(.001) \\

&$\gamma_{2}$AV&.903(.006)  .894(.000) 
&.958(.004)  1.07(.000)     
& .897(.006)  .283(.000) 
&.948(.003) .337(.000) \\

& $\rho_{p}$  &.977(.003)  1.25(.000)
& .994(.002)  1.49(.000)     
& .940(.004)  .395(.000)
& .971(.004)  .470(.000) \\

\hline

      & JEL$\gamma_{1}$ &.847(.007)  .532(.003)
      & .906(.005)  .611(.003)     
      & .890(.006)  .186(.001)
      & .942(.004) .209(.001) \\

      &AJEL$\gamma_{1}$  &.880(.007)  .572(.003)
      & .932(.004)  .666(.003)     
      & .896(.006)  .188(.001)
      & .945(.004)  .211(.001) \\

& $\gamma_{1}$J &.847(.008)  .717(.005) 
&.892(.005)  .854(.007)     
& .892(.005)  .221(.001) 
& .940(.004)  .263(.001) \\

& $\gamma_{1}$AV&.908(.005)  .699(.000) 
& .948(.004)  .833(.000)     
& .902(.005)  .221(.000) 
&.951(.004)  .264(.000) \\

$\rho=0.5$& JEL$\gamma_{2}$  &.848(.006)  .531(.002)
 &.906(.005)  .611(.003)     
 & .892(.005) .186(.001) 
 &.943(.004)  .208(.001) \\
 
& AJEL$\gamma_{2}$  &.881(.006)  .572(.003) 
&.933(.005)  .665(.003)     
& .897(.005)  .187(.001)
& .947(.004)  .211(.001) \\

 &$\gamma_{2}$J&.846(.006)  .712(.004)
 & .894(.006)  .849(.006)     
 & .894(.006)  .220(.001) 
 &.941(.004)  .262(.001) \\
 
& $\gamma_{2}$AV&.910(.006)  .699(.000) 
&.950(.004)  .833(.000)     
& .903(.005) .221(.000)
& .951(.004)  .264(.000) \\

 &$\rho_{p}$  &.956(.003)  .929(.000)
& .976(.003)  1.11(.000)     
& .936(.003)  .294(.000) 
 &.969(.003)  .350(.000) \\
 
\hline

       &JEL$\gamma_{1}$ &.841(.006)  .159(.001)   
       &  .895(.005)  .214(.003)            
        & .883(.004)  .048(.000) 
        &.938(.004) .062(.001) \\

    &  AJEL$\gamma_{1}$  &.871(.006)  .173(.002)  
     &         .917(.005)  .280(.003)             
     & .888(.004)  .049(.000)
     & .941(.004)  .064(.001) \\

& $\gamma_{1}$J &.827(.005)  .207(.002)         
&       .864(.006)  .245(.003)     
& .882(.005) .059(.000) 
&.929(.004)  .071(.000) \\

 &$\gamma_{1}$AV&.907(.004) .190(.000)    
 &         .932(.003)  .226(.000)     
 & .905(.006)  .060(.000) 
 &.945(.003)  .071(.000) \\

 $\rho=0.9$&JEL$\gamma_{2}$  &.842(.006)  .159(.002)      
 &           .894(.005)  .215(.003)            
 & .883(.005)  .048(.000) 
 &.938(.004)  .062(.000) \\
 
 &AJEL$\gamma_{2}$  &.871(.005)  .173(.002)     
 &       .917(.005)  .282(.003)     
 & .888(.005)   .049(.000) 
 & .941(.004)  .064(.001) \\
 
&  $\gamma_{2}$J&.828(.006)  .207(.002)     
&          .864(.006)  .245(.003)     
& .884(.004)  .059(.000) 
& .930(.004)  .071(.000) \\

&$\gamma_{2}$AV&.907(.005)  .190(.000)      
 &      .931(.003)  .226(.000)     
 & .906(.005)  .060(.000) 
 &.950(.003)  .071(.000) \\

& $\rho_{p}$  &.938(.004) .235(.000)      
   &       .957(.003)  .280(.000)     
   & .935(.004)  .074(.000) 
  &.966(.003)  .089(.000) \\

\hline\hline
\end{tabular}
\end{table}
%\end{sidewaystable}
%\renewcommand{\baselinestretch}{2}

Table \ref{tab2:covprho} lists the results under the bivariate $t(5)$ distribution.  As expected, the $\rho_P$ method performs poorly for heavy-tailed distributions. It suffers a serious over-coverage problem for all cases. For the $\gamma$AV method, the asymptotic variance (\ref{asyv}) is calculated by a Monte Carlo simulation with sample size of $10^8$. In this sense, we say $\gamma$AV to be a parametric method and it yields good coverage probabilities. For the two nonparametric methods, JEL and $\gamma$J,  both of them have slight under-coverage problems especially when $\rho$ is large and $n$ is small, but the JEL method produces better coverage probabilities and shorter confidence intervals than $\gamma$J. When the sample size is small ($n=20$) and $\rho$ is small ($\rho=0.1$),  JEL interval estimators are as short as half the length of  $\gamma$J  and $\gamma$AV interval estimators. Compared with $\gamma$AV methods, the  JEL method always has shorter confidence intervals.  Additionally the adjusted JEL methods improve the under-coverage problems. 
\section{JEL test for the equality of Gini correlations }

The two Gini correlations in (\ref{eqn:rho})  are not equal generally.  One sufficient condition for the equality of the two Gini correlations is that $X$ and $Y$ are exchangeable up to a linear transformation. That is,  there exist $a, b, c, $ and $d$ ($a, c >0$) such that $(X, Y)^T$ and $(aY+b, cX+d)^T$ are equally distributed. 
Particularly, if $(X, Y)^T$ are elliptically distributed with linear correlation parameter $\rho$, then $X$ and $Y$ are exchangeable up to a linear transformation. Hence we have $\gamma(X, Y)=\gamma(Y, X)$ and they are equal to $\rho$. More details are referred to (\cite{Schechtman87}, \cite{Yitzhaki13}).  Let $\Delta= \gamma(X,Y)-\gamma(Y,X)$. The hypotheses of interest are
\begin{align}\label{hypotest}
H_0: \Delta=0 \;\;  \text{vs}  \;\; H_a: \Delta \neq 0.
\end{align}
%If we have sufficient evidence to reject the null hypothesis, we are able to draw conclusions that the distribution of $X$ and $Y$ are not exchangeable up to a linear transformation, and hence are not exchangeable or not from the elliptically symmetric family. 
The objective of this section is to test the equality of the two Gini correlations via the JEL method.

\subsection{JEL test for the equality of the two Gini correlation}
For simplicity, we use $\gamma_1$ and $\gamma_2$ to denote $\gamma(X, Y)$  and $\gamma(Y, X)$, respectively.  Let $\bi \theta=(\Delta,\gamma_2)^T$ and we are interested in making inference of $\Delta$. Let $g_1((x_1,y_1), (x_2, y_2); \alpha)= h((x_1,y_1), (x_2,y_2); \alpha)$ and $g_2((x_1,y_1), (x_2, y_2); \alpha)= h((y_1,x_1),(y_2, x_2); \alpha)$, where $h$ is defined in (\ref{eqn:kernel}). We define a vector $U$-statistic type functional as
\begin{align*} 
\bi U_{n}(\bi \theta)={n \choose 2}^{-1}\sum_{1\le i<j \le n}\bi G((X_{i}, Y_{i}), (X_{j}, Y_{j});  \bi \theta),
\end{align*}
with kernels 
\begin{align*}
\bi G((x_1, y_1), (x_2, y_2);\bi \theta) =\left(\begin{array}{l}g_1((x_1, y_1), (x_2, y_2);\Delta+\gamma_2)\\
g_2((x_1, y_1), (x_2, y_2);\gamma_2) \end{array}\right).
\end{align*}
 It is easy to see that $\E \bi G((X_1, Y_1), (X_2, Y_2);\bi \theta)=\bi 0$. 
 
We do not apply the profile EL method since the computation of the profile EL could be very difficult even for equations without a $U$-structure involved. For our case, 
since $g_2(\cdot)$ function does not depend on $\Delta$, it enables us to estimate $\gamma_2$ by $\tilde{\gamma}_2$ that solves
\begin{align*}
\left( \begin{array}{ccc}
n  \\
2 \end{array} \right)^{-1}\sum_{1\le i<j\le n}g_2((X_i, Y_i), (X_j, Y_j);  \gamma_2)=0 .
\end{align*}
It is easy to check that $\tilde{\gamma}_2=\hat{\gamma}(Y,X)$, which can be easily computed with complexity $O(n\log n)$. We plug $\tilde{\gamma}_2$ in $g_1$ and conduct the JEL method for $\Delta$. More specifically,  
let 
\begin{align}\label{mn}
M_{n}(\Delta)=\frac{2}{n(n-1)} \sum_{1\leq i<j \leq n}g_1((X_i, Y_i), (X_j, Y_j);  \Delta+\tilde{\gamma}_2)
\end{align} 
and 
\begin{align*}
M_{n, -i}(\Delta)=\frac{2}{(n-1)(n-2)} \sum_{1\leq k<j \leq n, k \neq i}g_1((X_k, Y_k), (X_j, Y_j);  \Delta+\tilde{\gamma}_2).
\end{align*} 
Then the  jackknife pseudo samples are
\begin{align}\label{pseudo}
\hat{V}_{i}(\Delta)=nM_{n}(\Delta)-(n-1)M_{n, -i}(\Delta),\;\; i=1,...,n
\end{align}
and the jackknife empirical likelihood ratio at  $\Delta$ is 
\begin{align*}
R(\Delta)=\max \left\{ \prod_{i=1}^n (np_i): p_i \geq 0, i=1,...,n; \sum_{i=1}^n p_i=1; \sum_{i=1}^n p_i \hat{V}_{i}(\Delta)=0 \right\}.
\end{align*}
By the standard Lagrange multiplier method, we obtain the log empirical likelihood ratio  as 
\begin{align*}
\log R(\Delta)=-\sum_{i=1}^{n}\log \{1+\lambda \hat{V}_{i}(\Delta)\},
\end{align*}
where $\lambda=\lambda(\Delta)$ satisfies
\begin{align}\label{eqn:lambda}
f(\lambda)= \frac{1}{n} \sum_{i=1}^{n}\frac{\hat{V}_{i}(\Delta)}{1+\lambda \hat{V}_{i}(\Delta)}=0.
\end{align} 

Define 
$g((x, y);\Delta, \gamma_2)=\mathbb{E}h((x, y), (X_2, Y_2);\Delta+\gamma_2)$ and $\sigma^{2}_{g}(\Delta, \gamma_2)=\Var (g((X_1, Y_1);\Delta, \gamma_2)$. 
%and $\bar{\Delta}=\mbox{arg} \mbox{min}_{\Delta}l(\Delta),$ then
We have the following result.
\begin{theorem}\label{thtest}
If  $\E X^2_1<\infty$, $\E Y^2_1<\infty$ and $\sigma^{2}_{g}(\Delta, {\gamma_2})>0$, then 
$$-2 \log R(\Delta)\stackrel{d}{\rightarrow} \chi^{2}(1), \;\;\; \text{as n $\to$ $\infty$}.$$
\end{theorem}
A proof of Theorem \ref{thtest} needs to deal with an extra variation introduced by estimator $\tilde{\gamma}_2$, which  is given in the Appendix. 
\begin{remark}
Li $et$ $al.$ (\cite{Li2016}) established the Wilks' theorem for a general U-type profile empirical likelihood ratio under a strong condition that the kernel  $\bf G$ is uniformly bounded in both variables and parameters. Here we only assume the existence of the second moments of the kernel functions in the sample space. 
\end{remark}

%\begin{remark}
%Generally, Wilks' theorem does not hold for a plug-in empirical likelihood ratio. Instead, it converges in distribution to a weighted sum of independent chi-square random variables; see(\cite{Hjort09}). However, in our case, the nuisance parameter $\gamma(Y, X)$ can be estimated simply as a function of interested parameter $\Delta$, the limiting distribution only involves one chi-square random variable and its weight is proved to be one (\cite{Li2011}), which implies the standard Wilks' theorem since. 
%\end{remark}

\begin{remark}
The profile empirical likelihood ratio is usually computed through the ratio of the EL at the true value of the parameter and the EL at the maximal empirical likelihood estimate.  In our case, because $g_2$ is independent with $\Delta$ and $g_1$ is linear in $\Delta$ and $\gamma_2$, our JEL ratio does not involve  the  maximal empirical likelihood estimate, enjoying the computational ease property. 
\end{remark}

We can obtain  a $100(1-\alpha)\%$ jackknife empirical likelihood confidence interval for $\Delta$  as
\begin{align*}
I_{\alpha} = \{ \tilde{\Delta}: -2 \log \hat{R}(\tilde{\Delta})\le \chi^2_{1, 1-\alpha} \},
\end{align*}
where $\log \hat{R}(\tilde{\Delta})$ is the observed log-likelihood ratio at $\tilde \Delta$.
If $0 \notin I_{\alpha}$, we are able to reject $H_0$ at $\alpha$ significance level. 
For the hypothesis test (\ref{hypotest}), under the null hypothesis, the $p$-value can be calculated by
\begin{align*}
p\text{-value}=P(\chi^{2}_1>-2\log \hat{R}(0)),
\end{align*}
where $\chi^{2}_1$ is a random variable from a chi-square distribution with one degree of  freedom. For instance, under elliptical distributions we are able to compute $p$-values for the test.
On the other hand, in the case that the true parameter $\Delta_0\neq 0$, Theorem \ref{thtest} holds rather under $H_0$ but  under $H_a$,  and hence  the power of the test under $H_a: \Delta=\Delta_0 $ can be computed to be
\begin{align*}
\text{power}=1-P(\text{Accept} \ H_0 \arrowvert H_a)=1-P(-2\log R(\Delta_0)\leq \chi^{2}_{1, 1-\alpha})=P(-2\log R(\Delta_0) > \chi^{2}_{1, 1-\alpha}).
\end{align*} 
In the next, we consider simulations of two cases. One is for the case $\Delta=0$ and the other is for $\Delta \neq 0$.

\subsection{Empirical performance}
In the simulation, 3000 samples of two different sample sizes ($n=20, 200$) are drawn from bivariate normal and normal-lognormal distributions, that is,  $(X, Y)$ and $(X, \log Y)$  are drawn from $N(\bi 0, \bi \Sigma)$   with the scatter matrix $\bf \Sigma=\begin{pmatrix}4 \;\;\;&2\rho\\ 2\rho\;\;\;&1\end{pmatrix}$.
Under the bivariate normal distribution, $\gamma(X ,Y)=\gamma(Y, X)=\rho$, the null hypothesis, $\Delta=0$, is true and  $p$ values are provided in Table \ref{tab4:covprho} along with averages and standard deviations of the coverage probabilities. Under the mixture of normal and lognormal distribution, $\gamma(X, Y)=\rho$ and  $\gamma(Y, X)=\frac{2\Phi(\rho \sigma_2/\sqrt{2})-1}{2\Phi( \sigma_2/\sqrt{2})-1}$ is not equal to $\gamma(X,Y)$ for $\rho \neq 0$. Thus powers of the test are presented in Table \ref{tab5:covprho}. 
%For each simulated data set, $90\%$ and $95\%$ confidence intervals are calculated. The procedure is repeated 30 times. The average coverage probabilities and $p$-values or powers  are calculated using different methods. 
% The JEL methods are denoted by  `JELdelta', and the adjusted JEL method (`AJELdelta') is also considered. The method based on the limiting normal distribution with a jackknife asymptotic variance estimate is denoted as `deltaJ'. 

%$U$-statistics theorem,  $\hat{\Delta}$, will, after appropriate normalization, have a limiting normal distribution. However, the variance formula is complicated. Schechtman $et$ $al.$ (\cite{Schechtman07}) recommended  an estimation of the variance based on the jackknife method. 
%$\hat {\Delta}_{(-i)}$ is the jackknife values. Then 
%\begin{align}\label{jelv}
%\hat{V}(\hat{\Delta}) =\frac{n-1}{n}\sum_{i=1}^n ({\hat{\Delta}}_{(-i)} -\bar{\hat{\Delta}}_{(\cdot)})^2,
%\end{align}
%where $\bar{\hat{\Delta}}_{(\cdot)} =1/n \sum_{i=1}^n {\hat{\Delta}}_{(-i)}$. 
%We use `deltaJ' to denote the the above method.

\begin{table}[H]
%\center
%\begin{sidewaystable}[thb]
\center
\tiny
\scriptsize
%\footnotesize
\caption{Coverage probabilities (standard deviations) of  interval estimators of $\Delta$ and $p$-values (standard deviations) of the test for $\Delta=0$ under bivariate normal distributions. }
\label{tab4:covprho}
\begin{tabular}{ll|c   c|c  c }
\hline \hline
 && \multicolumn{2}{c|}{$n=20$}&
 \multicolumn{2}{c}{$n=200$}\\
$\Delta_0(\rho)$&{Method}&{$1-\alpha=0.90$}   & {$1-\alpha=0.95$}& {$1-\alpha=0.90$}   & {$1-\alpha=0.95$}   \\
$$&&CovProb  P-Value& CovProb  P-Value&CovProb  P-Value &CovProb  P-Value \\
\hline
$\rho=0.1$&&&&&\\
        & JELdelta&.918(.005)  .515(.004) 
        &.962(.004)  .515(.005)     
        & .905(.005)  .505(.005) 
        &.952(.003)  .505(.006) \\

$\Delta_0=0$&AJELdelta   &.965(.004)  .565(.004) 
&.991(.002)  .565(.005)     
& .909(.005)  .509(.005) 
&.955(.003)   .509(.005) \\

     & deltaJ  &.964(.004)  .747(.005) 
     &.986(.002)   .745(.005)     
     & .914(.005)   .723(.005)
     & .958(.003)   .724(.006) \\

\hline
$\rho=0.5$&&&&&\\

       & JELdelta&.941(.004)  .537(.005) 
       &.975(.003)  .537(.004)     
       & .912(.005)  .509(.004) 
       &.958(.004)  .511(.005) \\

$\Delta_0=0$&AJELdelta   &.979(.002)  .586(.005) 
&.996(.001)  .586(.004)     
& .916(.005)  .514(.004) 
&.961(.004)  .515(.005) \\

  &    deltaJ  &.978(.003)  .750(.006)
  & .993(.002)  .749(.004)     
  & .925(.005)  .726(.004) 
  &.967(.004)  .727(.006) \\
  \hline
$\rho=0.9$&&&&&\\

&         JELdelta&.971(.004)  .582(.005) 
&      .991(.002)  .580(.004)       
& .962(.004)   .564(.004)
 &.987(.002)  .567(.005) \\

$\Delta_0=0$&AJELdelta   &.994(.002)  .629(.004)    
&       .999(.000)   .627(.004)     
& .964(.004)   .568(.004)
 &.989(.002)  .571(.005) \\

&      deltaJ  &.993(.002)  .756(.004)      
&    .999(.001)  .754(.005)     
& .971(.003)  .743(.004) 
&.992(.002)  .742(.005) \\
\hline\hline
\end{tabular}
\end{table}

\begin{table}[htbp]
%\center
%\begin{sidewaystable}[thb]
\center
\tiny
\scriptsize
%\footnotesize
\caption{Coverage probabilities (standard deviations) of  interval estimators of $\Delta$ and powers (standard deviations) of the test for $\Delta =\Delta_0$ under the normal-lognormal distributions.  }
\label{tab5:covprho}
\begin{tabular}{ll|c   c|c  c }
\hline \hline
 && \multicolumn{2}{c|}{$n=20$}&
 \multicolumn{2}{c}{$n=200$}\\
$\Delta_0, \rho$&{Method}&{$1-\alpha=0.90$}   & {$1-\alpha=0.95$}& {$1-\alpha=0.90$}   & {$1-\alpha=0.95$}   \\
&&CovProb  Power &CovProb  Power&CovProb  Power&CovProb  Power\\
\hline
$\rho=0.1$&&&&&\\
       &  JELdelta&.857(.007)  .256(.007) 
       &.921(.005)  .196(.008)    
        & .873(.006)  .192(.008) 
        &.930(.005) .154(.006) \\

$\Delta_0=-.008$&AJELdelta   &.896(.006)  .226(.007) 
&.950(.004)  .155(.006)     
& .879(.006)  .189(.008) 
&.933(.005)  .151(.006) \\

    &  deltaJ  &.938(.004)  .062(.004) 
    &.970(.003)  .031(.003)     
    & .893(.006) .110(.006)  
    &.945(.004)  .056(.004) \\

\hline
$\rho=0.5$&&&&&\\
         &JELdelta&.858(.007)  .213(.006) 
         &.917(.005)  .161(.007)     
         & .876(.005)  .235(.008)   
         & .932(.005)  .171(.006) \\

$\Delta_0=-.031$&AJELdelta   &.895(.006)  .184(.005)  
&.945(.004)  .127(.006)     
& .881(.005)  .230(.008)  
&.936(.005)  .166(.006) \\

   &   deltaJ  &.947(.004)  .052(.004) 
   &.978(.003)  .021(.002)     
   & .897(.005)  .164(.007)  
   &.949(.004) .094(.004) \\

\hline
$\rho=0.9$&&&&&\\
    &     JELdelta&.865(.005)  .144(.007) 
    &      .916(.004)  .093(.005)       
    & .884(.005)  .370(.009)  
    &.939(.004)  .268(.010) \\

$\Delta_0=-.014$& AJELdelta   &.898(.005)  .114(.006)   
&        .941(.004)  .066(.004)     
& .889(.005)  .362(.008) 
&.943(.004)  .261(.010) \\

   &   deltaJ  &.976(.003)  .031(.004)   
   &       .993(.002)  .008(.001)     
   & .907(.005)   .293(.008) 
   &.955(.004)   .187(.008) \\
\hline\hline
\end{tabular}
\end{table}

%Simulation results for bivariate normal distributions are reported in Table \ref{tab4:covprho} and for the normal-lognormal distributions in Table \ref{tab5:covprho} . 

Under bivariate normal distributions, both JEL and deltaJ methods have over coverage probability problems especially for large $\rho$ under small sample size.  The JEL method performs relatively better than the deltaJ method for all sample size and all $\rho$ values. The $p$-values in Table \ref{tab4:covprho}  are all greater than 0.5, which indicates that we cannot reject $H_0$ under a bivariate normal distribution. This implies no evidence to reject the exchangeability up to a linear transformation, which is a correct decision under bivariate normal distributions.

From Table \ref{tab5:covprho}, under normal-lognormal distributions we can observe that the powers of  the test for all the listed methods are not high. This can be explained by the fact that the true values $\Delta_0$ are very close to 0, making the procedures difficult to reject $H_0$. However, the JEL method is more efficient with higher powers for all sample sizes than deltaJ. 
 Among all the approaches in Table \ref{tab5:covprho}, deltaJ method produces good coverage probabilities when the sample size is large but have serious over-coverage problems when $n=20$ and $\rho$ is large. This is due to one characteristic of the lognormal distribution. The bias and variance of the sample correlation may be quite significant especially when the correlation coefficient $\rho$ is not close to zero (\cite{Witting95}). On the other hand, the JEL method has under-coverage problems when the sample size is small but these problems have been corrected effectively by the adjusted JEL method.

\section{JEL for independent data}
%Comparing the Gini correlations from independent data sets is of considerable interest. We develop the JEL method to do the comparison in this section. 
Let $\mathcal{Z}_1 =\{{\bf Z}^{(1)}_{1},...,{\bf Z}^{(1)}_{n_1}\}$ and $\mathcal{Z}_2 =\{{\bf Z}^{(2)}_{1},...,{\bf Z}^{(2)}_{n_2}\}$, where ${\bf Z}^{(j)}_i =(X^{(j)}_i,Y^{(j)}_i)^T$ for $j=1, 2$, be independent samples from distributions $\bi H(x, y)$ and $\bi M(x, y)$ with sample size $n_1$ and $n_2$, respectively. Let $\gamma^{(1)}_{1}$,  $\gamma^{(1)}_{2}$, $\gamma^{(2)}_{1}$ and  $\gamma^{(2)}_{2}$ denote the Gini correlations between $X$ and $Y$, $Y$ and $X$ for these two distributions, respectively.  Let $\delta_1=\gamma^{(1)}_{1}-\gamma^{(2)}_{1}$ and $\delta_2=\gamma^{(1)}_{2}-\gamma^{(2)}_{2}$, the hypotheses of interest are
\begin{align}\label{hypotest2}
&H_0:\begin{pmatrix}
\delta_1\\
\delta_2
\end{pmatrix}
=\begin{pmatrix}
0\\0
\end{pmatrix} \;\;  \text{vs}  \;\; H_a: \begin{pmatrix}
\delta_1\\
\delta_2
\end{pmatrix}
\neq \begin{pmatrix}
0\\0
\end{pmatrix}. 
\end{align}

Our aim for this section is to derive a JEL method to test the above statements.  
%construct confidence intervals  for the differences $\delta_1$ and $\delta_2$ and to perform the above hypothesis tests by utilizing the JEL method.  

\subsection{JEL for Gini correlation differences for independent data}

Due to independence of ${\bf Z}_i^{(j)}$ for $i=1,2$ and $j=1,2$,  we have

\begin{align*}
&\delta_1 =  
 \frac{\E [h_1({\bf Z}^{(1)}_{1},{\bf Z}^{(1)}_{2})h_2({\bf Z}^{(2)}_{1},{\bf Z}^{(2)}_{2}) - h_2({\bf Z}^{(1)}_{1},{\bf Z}^{(1)}_{2})h_1({\bf Z}^{(2)}_{1},{\bf Z}^{(2)}_{2})] }{\E h_2( {\bf Z}_1^{(1)}, {\bf Z}_2^{(1)}) h_2( {\bf Z}_1^{(2)}, {\bf Z}_2^{(2)})}
\end{align*}
and 
\begin{align*}
&\delta_2 =  
 \frac{\E [h'_1({\bf Z}^{(1)}_{1},{\bf Z}^{(1)}_{2})h'_2({\bf Z}^{(2)}_{1},{\bf Z}^{(2)}_{2}) - h'_2({\bf Z}^{(1)}_{1},{\bf Z}^{(1)}_{2})h'_1({\bf Z}^{(2)}_{1},{\bf Z}^{(2)}_{2})] }{\E h'_2( {\bf Z}_1^{(1)}, {\bf Z}_2^{(1)}) h'_2( {\bf Z}_1^{(2)}, {\bf Z}_2^{(2)})},
\end{align*}
where $h'_i(\bi z_1, \bi z_2)=h_i((y_1, x_1), (y_2, x_2))$, $i=1, 2.$
This motivates us to define a two-sample $U$-statistic type functional  as
\begin{align}\label{U22}
&\bi U_{n_1, n_2}(\delta_1, \delta_2)={n_1 \choose 2}^{-1} {n_2 \choose 2}^{-1}\sum_{1\le i_1<i_2 \le n_1} \sum_{1\le j_1<j_2 \le n_2}\bi H ( {\bf Z}^{(1)}_{i_1},{\bf Z}^{(1)}_{i_2}, {\bf Z}^{(2)}_{j_1},{\bf Z}^{(2)}_{j_2};  \delta_1, \delta_2)\\\nonumber
&=:\bi T({\bf Z}^{(1)}_{1},...,{\bf Z}^{(1)}_{n_1},{\bf Z}^{(2)}_{1},...,{\bf Z}^{(2)}_{n_2};\delta_1, \delta_2)
\end{align}
with 
\begin{multline*}
\bi H ( {\bf z}^{(1)}_{1},{\bf z}^{(1)}_{2}, {\bf z}^{(2)}_{1},{\bf z}^{(2)}_{2};  \delta_1, \delta_2)=\\
\begin{pmatrix} 
h_2({\bf z}^{(1)}_{1},{\bf z}^{(1)}_{2})h_2({\bf z}^{(2)}_{1},{\bf z}^{(2)}_{2})\delta_1 -h_1({\bf z}^{(1)}_{1},{\bf z}^{(1)}_{2})h_2({\bf z}^{(2)}_{1},{\bf z}^{(2)}_{2})
+h_2({\bf z}^{(1)}_{1},{\bf z}^{(1)}_{2})h_1({\bf z}^{(2)}_{1},{\bf z}^{(2)}_{2})\\
h'_2({\bf z}^{(1)}_{1},{\bf z}^{(1)}_{2})h'_2({\bf z}^{(2)}_{1},{\bf z}^{(2)}_{2})\delta_2 -h'_1({\bf z}^{(1)}_{1},{\bf z}^{(1)}_{2})h'_2({\bf z}^{(2)}_{1},{\bf z}^{(2)}_{2})
+h'_2({\bf z}^{(1)}_{1},{\bf z}^{(1)}_{2})h'_1({\bf z}^{(2)}_{1},{\bf z}^{(2)}_{2})
\end{pmatrix}
\end{multline*}
It is easy to check that 
$\begin{pmatrix}
\delta_1\\
\delta_2
\end{pmatrix}$
is the solution of $\mathbb{E}\bi H ( {\bf Z}^{(1)}_{i_1},{\bf Z}^{(1)}_{i_2}, {\bf Z}^{(2)}_{j_1},{\bf Z}^{(2)}_{j_2};  {\delta_1}, \delta_2)=0$. To apply the JEL to $\bi U_{n_1, n_2}(\delta_1, \delta_2)$, let $n=n_1+n_2$ and 
$$\bi Q_i = \begin{cases} \bi Z^{(1)}_{i}, &  i=1, ..., n_1; \\ \bi Z^{(2)}_{(i-n_1)}, &i=n_1+1,...,n. \end{cases}$$ 
Then we consider $\mathcal{Q}=\{\bi Q_1, \bi Q_2, ..., \bi Q_n\}$ as a new sample,  and 
by (\ref{U22}), we have
\begin{align*}
\bi U_{n_1, n_2}(\delta_1, \delta_2)=\bi T({\bf Z}^{(1)}_{1},...,{\bf Z}^{(1)}_{n_1},{\bf Z}^{(2)}_{1},...,{\bf Z}^{(2)}_{n_2};\delta_1, \delta_2)=\bi T(\bi Q_1, \bi Q_2, ..., \bi Q_n;\delta_1, \delta_2).
\end{align*}
The corresponding jackknife pseudo-values are given by
\begin{align*}
 \hat{\bi V}_{i}(\delta_1, \delta_2)=n\bi T_{n}(\delta_1, \delta_2)-(n-1)\bi T^{(-i)}_{n-1}(\delta_1, \delta_2).
\end{align*}
 It can be easily shown that  $\mathbb{E}[\hat{\bi V}_{i}(\delta_1, \delta_2)]=\bi 0$ and
\begin{align*}
\bi U_{n_1, n_2}(\delta_1, \delta_2)=\frac{1}{n}\sum_{i=1}^n\hat{\bi V}_{i}(\delta_1, \delta_2).
\end{align*}

%Then we apply the regular EL method to the jackknife pseudo values $\hat{V}_{i}(\Delta_1)$.
Following the lines as Section 2 did, 
we have the JEL  ratio at $(\delta_1, \delta_2)$ to be
\begin{align*} 
R(\delta_1, \delta_2)=\max \left \{\prod_{i=1}^{n}(np_i): p_i \ge 0, i=1,...,n;  \sum_{i=1}^n p_i=1; \sum_{i=1}^n p_i \hat{\bi V}_{i}(\delta_1, \delta_2)=\bi 0 \right \}.
\end{align*}
Utilizing the standard Lagrange multiplier technique, the jackknife empirical log-likelihood ratio at $(\delta_1, \delta_2)$ is
\begin{align*}
\log R(\delta_1, \delta_2)=-\sum_{i=1}^{n}\log [1+\bi \lambda^T \hat{\bi V}_{i}(\delta_1, \delta_2) ],
\end{align*}
where $\bi \lambda=\bi \lambda(\delta_1, \delta_2)$ satisfies 
\begin{align*}
\frac{1}{n}\sum_{i=1}^n \frac{ \hat{\bi V}_{i}(\delta_1, \delta_2)}{1+\bi \lambda^T \hat{\bi V}_{i}(\delta_1, \delta_2)}=\bi 0.
\end{align*}
Define 
\begin{align*}
&\bi g_{1, 0}((x, y);\delta_1, \delta_2)=\E \bi H((x, y), (X^{(1)}_{2}, Y^{(1)}_{2}),  (X^{(2)}_{1}, Y^{(2)}_{1}), (X^{(2)}_{2}, Y^{(2)}_{2});\delta_1, \delta_2),\\
&\bi \Sigma_{1,0}^{2}(\delta_1,  \delta_2)=\mbox{Cov} (\bi g_{1,0} ((X^{(1)}_{1}, Y^{(1)}_{1}); \delta_1,  \delta_2)),\\
&\bi g_{0, 1}((x, y);\delta_1,  \delta_2)=\E \bi H((X^{(1)}_{1}, Y^{(1)}_{1}), (X^{(1)}_{2}, Y^{(1)}_{2}), (x, y), (X^{(2)}_{2}, Y^{(2)}_{2});\delta_1,  \delta_2),\\
&\bi \Sigma_{0,1}^{2}(\delta_1)=\mbox{Cov} (\bi g_{0, 1}((X^{(2)}_{1}, Y^{(2)}_{1});\delta_1,  \delta_2)).
\end{align*}
We have the following Wilks' theorem.
\begin{theorem}\label{wilkrhodif}
Assume 

1. $\E[X^{(j)}_1]^2<\infty$, $\E [Y^{(j)}_1]^2<\infty$, $j=1, 2$;

2. $\bi \Sigma_{1,0}^{2}(\delta_1, \delta_2)$ and $\bi \Sigma_{0,1}^{2}(\delta_1, \delta_2)$ are positive definite;

3. $n_1/n_2 \to r$, where $0<r<\infty$.
 Then we have 
$$ -2 \log  R(\delta_1, \delta_2)\stackrel{d}{\rightarrow} \chi^{2}_2, \;\;\;\text{as $n \to \infty$}.$$
\end{theorem}
A  $100(1-\alpha)\%$ jackknife empirical likelihood joint confidence region for $(\delta_1, \delta_2)$  can be constructed  as
\begin{align}\label{con_reg}
I_{\alpha} = \{ (\tilde{\delta}_1, \tilde{\delta}_2): -2 \log \hat{ R}(\tilde{\delta}_1, \tilde{\delta}_2)\le \chi^2_{2, 1-\alpha} \},
\end{align}
where $\log \hat{ R}(\tilde{\delta}_1, \tilde{\delta}_2)$ is the observed log-likelihood ratio at $(\tilde{\delta}_1, \tilde{\delta}_2) $.
%where $\chi^2_{1, 1-\alpha}$ denotes the $(1-\alpha)$ quantile of the chi-square distribution with one degree of freedom.
The null hypothesis of test (\ref{hypotest2}) is rejected at $\alpha$ significance level if $\bi 0 \notin I_{\alpha}$.
%\begin{remark}
%For $\delta_2$, similar results can be obtained.
%\end{remark}

\subsection{Empirical performance}
Simulation studies are conducted for two independent samples with either equal  ($n_1=n_2=20$) or unequal sample sizes ($n_1=150, n_2=200$).  We generate 3000 independent samples of $(X, Y)$ and $(T, \log W)$ from ${N( \bi 0, \bi \Sigma)}$  with the scatter matrix $\bi\Sigma=\begin{pmatrix}1 \;\;\;&4\rho\\ 4\rho\;\;\;&16\end{pmatrix}$ and consider $\delta_{1}=\gamma(X,Y)-\gamma(T, W)$  and $\delta_{2}=\gamma(Y, X)-\gamma(W, T)$.  
%We test two hypothesis: $H^1_0:\delta_{1}=0$ and $H^{2}_0: \delta_{2}=0$.   

As mentioned in Section 3,  $\delta_1:=\gamma(X,Y)-\gamma(T, W)=0$ and $\delta_2:=\gamma(Y, X)- \gamma(W, T)\neq 0$. For this simultaneous test,  coverage probabilities and $p$-values for the test (\ref{hypotest2}) are provided for $\rho=0.1, 0.5, 0.9$, in Table \ref{indepairdelta},  respectively.

\begin{table}[!htbp]
\center
%\begin{sidewaystable}[thb]
%\center

\scriptsize
\footnotesize
\caption{$p$-values (standard deviation) of the test (\ref{hypotest2}) at significance level $\alpha=0.10$. }
\label{indepairdelta}
\begin{tabular}{c   c|c c  }
\hline \hline
%\begin{tabular}{l|c   c|c  c }
%\hline \hline
 &{Method}& \multicolumn{1}{c}{$n_1=n_2=20$}&
 \multicolumn{1}{c}{$n_1=150, \;n_2=200$}\\
\hline

    $\rho=0.1$  &   JELdelta&  .46848(.0042) 
        
      &   .3562(.0038) \\

$\delta_{20}=-.1237$&AJELdelta  & .4985(.0041)  
  
&   .3617(.0038)
 \\

      \hline

 $\rho=0.5$    &   JELdelta&  .4696(.0053) 
        
      &  .2819(.0044) 
      \\

$ \delta_{20}=-.3467$&AJELdelta  &  .5001(.0053)  
  
&   .2869(.0044)
 \\

      \hline
    
    $\rho=0.9$ &   JELdelta&  .4701(.0037) 
        
      &   .1432(.0035) 
      \\

$ \delta_{20}=-.0937$&AJELdelta  &  .5020(.0037)  
  
&  .1474(.0035)
 \\

\hline\hline
\end{tabular}
\end{table}
From Table  \ref{indepairdelta}, we have observed that all the listed $p$-values are strictly greater than the nominal level $\alpha$, and we are not able to reject the null hypothesis which in fact is false. It may be due to the reason that the joint confidence regions have much smaller areas than the confidence regions constructed by the marginal confidence intervals, which can be seen in Figure 1 of next section. However, the large $p$-values decrease greatly for the large sample size ($n_1=150, n_2=200$).

\begin{remark}
Following a similar procedure, we can develop JEL approach to do a marginal test $H_0: \delta_1=0$ vs. $H_a: \delta_1 \neq 0$ or $H_0: \delta_2=0$ vs. $H_a: \delta_2 \neq 0$. Simulation studies (not presented here)  show that the JEL method performs well with good coverage probabilities and has a higher power than the method using jackknife method to estimate the asymptotic variance. 
\end{remark}

\section{Application}
For the purpose of illustration, we apply the JEL method to the banknote authentication data which is available in UCI Machine Learning Repository (\cite{Lich2013}).  The data set consists of 1372 samples with 762 samples of them  from the Genuine class denoted as Gdata and 610 from the Forgery class denoted as Fdata. Four features are recorded  from each sample: variance of wavelet transformed image (VW), skewness of wavelet transformed image (SW), kurtosis of wavelet transformed image (KW) and entropy of image (EI). One can refer to Lohweg $et$ $al.$ (\cite{Lohweg2013}) for more descriptions of the data.
Here we shall study the correlations of two pairs, namely VW and SW, SW and KW.

\begin{table}[!htbp]
\center
\scriptsize
\caption{Point estimates and $90\%$ confidence intervals for $\gamma_1, \gamma_2$ and $\Delta$ for two pairs of (VW, SW) and (SW,  KW), respectively . }
\label{tab6:test}
\begin{tabular}{l|c   c|c  c }
\hline \hline
 & \multicolumn{2}{c|}{Fdata}&
 \multicolumn{2}{c}{Gdata}\\
{Method}&{Point estimate}&{Confidence interval}  &{Point estimate}&{Confidence interval}     \\
\hline
 (VW, SW) \\ \hline
JEL&$\hat{\gamma}_{1}=.0471$
         & (-.0197, 0.1329) &$\hat{\gamma}_{1}=-.2459$
         &(-.2936, -.1828)\\

AJEL & & (-.0197, 0.1334)& &(-.2936, -.1826) \\

      VJ  && (-.0407, 0.1349) & &(-.3087, -.1831)\\
\hline
JEL&$\hat{\gamma}_{2}=.1595$
         & (0.0992, 0.2365)&$\hat{\gamma}_{2}=-.1916$
         & (-.2370, -.1317)\\

AJEL & & (0.0992, 0.2369)&& (-.2370, -.1315) \\

      VJ & & (0.0803, 0.2388) && (-.2513, -.1319)\\
\hline
JEL& $\hat{\Delta}=-.1124$
         &(-.1324, -.0865)& $\hat{\Delta}=-.0543$
         &(-.0685, -.0358)\\

AJEL & & (-.1324, -.0864) &&(-.0685, -.0357) \\

      VJ & & (-.1387, -.0861) &&(-.0730, -.0355) \\
\hline\hline

(SW,  KW)\\ \hline
JEL&$\hat{\gamma}_{1}=-.8436$
         & (-.8632, -.8169) &  $\hat{\gamma}_{1}=-.7638$&(-.7867, -.7320)\\

AJEL & & (-.8632, -.8167) & &(-.7867, -.7318)\\

      VJ  & & (-.8694, -.8178) & &(-.7939, -.7337)\\
\hline
JEL&$\hat{\gamma}_{2}=-.8910$
         & (-.9023, -.8755)&$\hat{\gamma}_{2}=-.7525$
         & (-.7752, -.7206)\\

AJEL & & (-.9023, -.8754)&& (-.7752, -.7205) \\

      VJ & & (-.9058, -.8762)&& (-.7823, -.7228) \\
\hline
JEL& $\hat{\Delta}=.0474$
         &(0.0363, 0.0628)& $\hat{\Delta}=-.0113$
         &(-.1840, -.0017)\\

AJEL & & (0.0363, 0.0629) &&(-.1840, -.0017) \\

      VJ & & (0.0327, 0.0620) &&(-.0206, -.0019) \\
\hline\hline
\end{tabular}
\end{table}
The point estimates and interval estimates for two Gini correlations and their difference are reported in Table \ref{tab6:test}.  Point estimators are quite different for the Genuine and Forgery classes. In Gdata, VW and SW have negative Gini correlation estimates and their confidence intervals contain only negatives, while estimates of  $\gamma_2$ in Fdata are positive.  All the confidence intervals for $\Delta$ in both classes do not contain 0, which indicates that either the pair variables VW and SW or the pair variables SW and KW  are not exchangeable. Also it means a sufficient evidence at the significance level 0.1 to reject that (VW, SW) or (SW, KW) has a elliptical distribution in Forgery class or Genuine class. In addition, compared with the VJ method (using jackknife method to estimate the asymptotic variance) , the JEL method provides shorter confidence intervals.

Denote $\delta_1$ and $\delta_2$ as the differences of Gini correlations between Gdata and Fdata about pairs of variables $(VW, SW)$ or $(SW, KW)$. We conduct simultaneous inference on $(\delta_1, \delta_2)$ . Using the JEL method of (\ref{con_reg}) in Section 4, a $90\%$ joint confidence region is obtained and plotted in Figure 1, along with two rectangular confidence regions formed by marginal inferences. The simultaneous ellipse-shaped regions clearly have smaller areas than the marginal ones. This is because the simultaneous JEL takes dependent information of the marginal JEL's of $\delta_1$ and $\delta_2$, and hence produces smaller confidence regions. Note that $(0, 0)$ falls outside all confidence regions. Therefore, we have evidence to reject the null hypothesis of $\delta_1=\delta_2=0$.

\begin{figure}[!htbp]
\centering
\label{con_reg}
\begin{tabular}{cc}
\includegraphics[width=3in,height=3in]{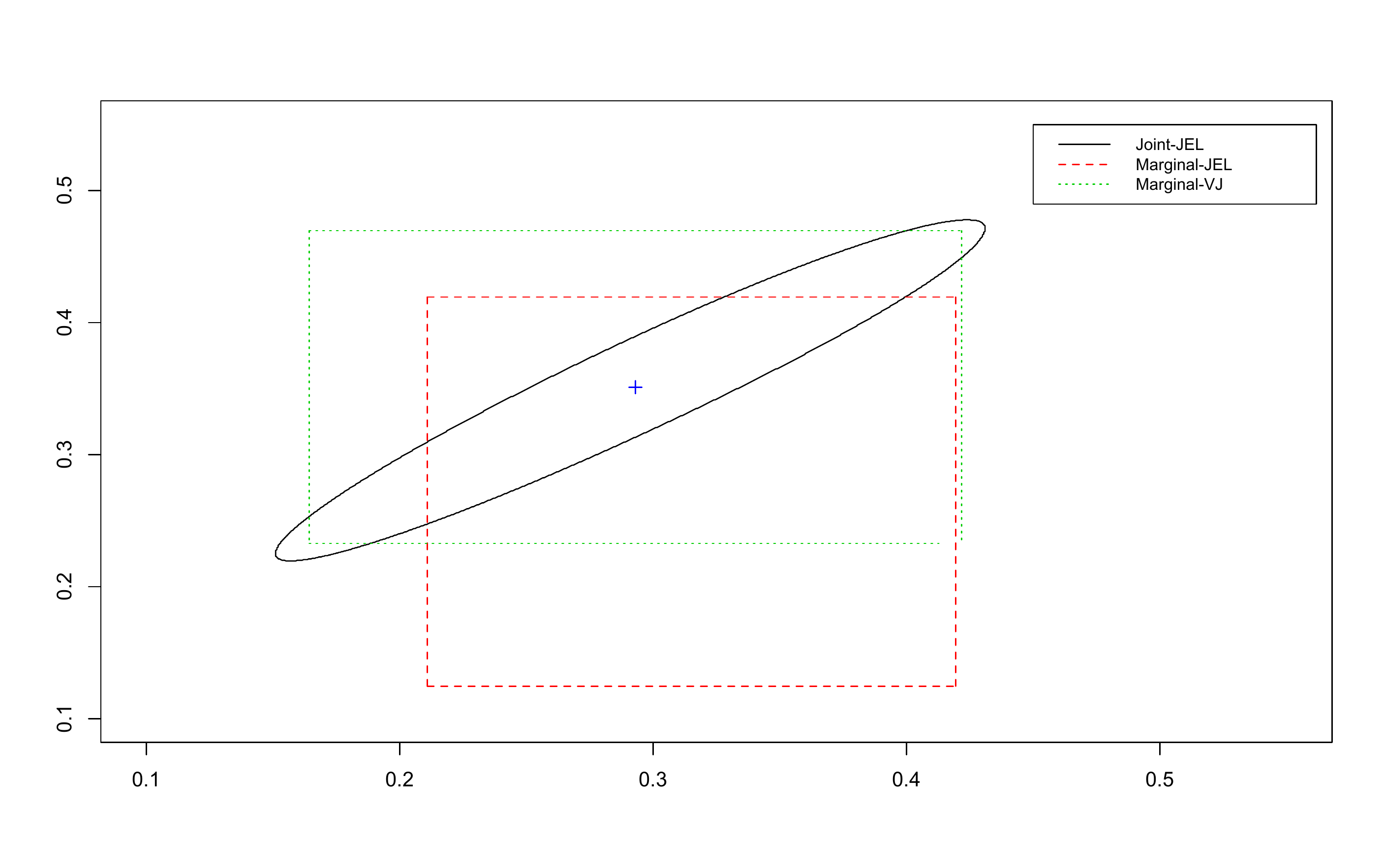} \vspace{-0.1in}&
\includegraphics[width=3in,height=3in]{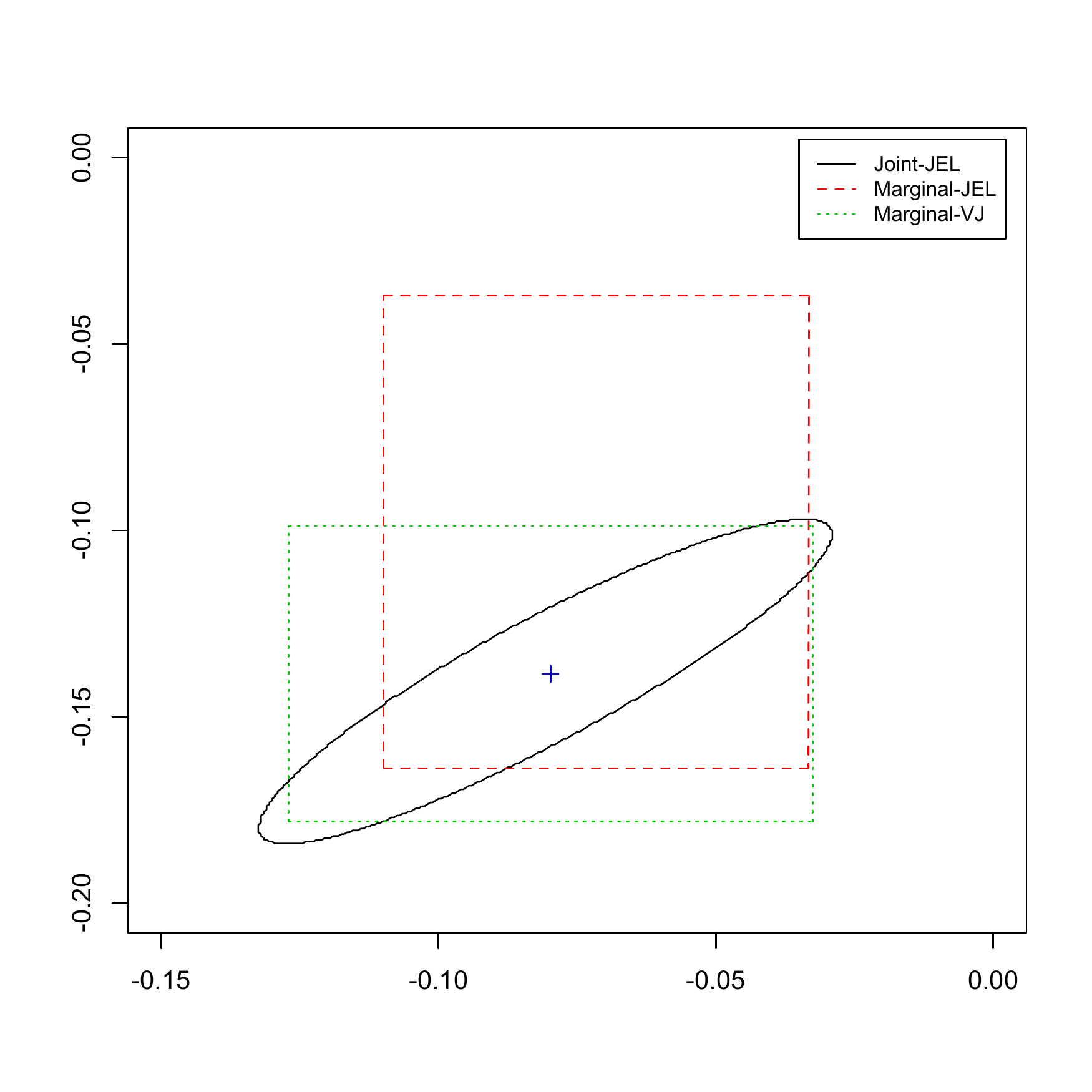} \vspace{-0.1in}\\
(a) (VW, SW) \vspace{-0.1in}&(b) (SW, KW)\vspace{0.05in}\\
\end{tabular}
\caption{Confidence regions of Gini differences $(\delta_1, \delta_2)$ about different pairs of variables. `+' indicates the point estimate.}
\label{fig:IF}
\end{figure}

\section{Conclusion}

In this paper, we have explored JEL methods for the Gini correlations. Three cases, namely Gini correlation, difference of two types of Gini correlation and difference of Gini correlations of two samples,  have been studied. For each case,  a novel $U$-structured equation has been defined and the associated jackknife empirical likelihood has been developed. By checking conditions and establishing the necessary lemmas, we have proved  the Wilks' theorem for each of the cases. Hence, the standard chi-square distribution is used to construct confidence intervals and to conduct hypothesis testings without estimating the asymptotic variance. Numerical studies confirm the advantages of the proposed method under a variety of situations. 

One of important contributions of this paper is to develop the JEL procedure with  $U$-structured estimating functions for inferences on the difference of two types of Gini correlations.  The procedure involves the parameter of interest and a nuisance parameter.  The simple plug-in jackknife empirical likelihood ratio holds the Wilks' theorem under the assumption of finite second moment of the distribution. It enjoys the theoretical and computational advantages due to its special form. For a general $U$-type profile empirical likelihood,  Li {\em et al. }\cite{Li2016} obtained its Wilks' theorem under strong conditions and they did not discuss the implementation of its computation. Continuations of this work could take the following directions: 
\begin{itemize}
\item Establish the Wilks' theorem for the $U$-type profile empirical likelihood ratio under weaker conditions. For example, one might want to relax the boundedness condition for the kernel functionals to a  moment finiteness condition. 
\item Reduce the computation of  $U$-type profile empirical likelihood. Li {\em et al.} \cite{Li2011} and Peng \cite{Peng2012} considered procedures based on a jackknife plug-in empirical likelihood to save the computation time. We may develop similar procedures to deal with $U$-structured empirical likelihood.  
\end{itemize} 

\section{Appendix}
In this section $a=O(b)$ means that $a/b$ is bounded and $a=o(b)$ means that $a/b \to 0$.

 \medskip
\noindent
\textbf{Proof of Theorem \ref{wilkrho}}
\begin{lemma}\label{lemma}
If $\E X^2_1<\infty$ and  $\E Y^2_1<\infty$.  Then $\E h^2((X_1, Y_1), (X_2, Y_2);\gamma)<\infty$ for $\gamma \in [-1, 1]$.
\end{lemma}
\textbf{Proof.}
\begin{align*}
&\E [h_1((X_1, Y_1), (X_2, Y_2))]^2=\frac{1}{16} \E [(X_1-X_2)^2  I (Y_1>Y_2) +(X_2-X_1)^2 I(Y_2>Y_1)]<\infty,\\
&\E [h_2((X_1, Y_1), (X_2, Y_2))]^2=\frac{1}{16}\E |X_1-X_2|^2 <\infty.
\end{align*}
Therefore,
\begin{align*}
&\E [h((X_1, Y_1), (X_2, Y_2);\gamma)]^2\leq 2  \gamma^2\E [ h_2((X_1, Y_1), (X_2, Y_2))]^2+2 \E [h_1((X_1, Y_1), (X_2, Y_2))]^2<\infty.
\end{align*}$\hfill{\Box}$

Although $U_n(\gamma)$ in (\ref{eqn: U_ngamma}) is not a $U$-statistic,  based on Lemma \ref{lemma}, we have similar results as Lemma A.1, A.2, A.3, A.4 and Corollary A.1 of \cite{Jing2009} by replacing $U_n-\theta$, $\sigma_g$ and $\hat{V}_i-\theta$ in \cite{Jing2009} with $U_n(\gamma)$, $ \sigma_g(\gamma)$ and $\hat{V}_i(\gamma)$, respectively. Then the  proof of Theorem \ref{wilkrho} is similar to  the proof of Theorem 1 in \cite{Jing2009} by using these results, and hence is skipped.

\bigskip
\bigskip

\medskip
\noindent
\textbf{Proof of Theorem \ref{thtest}}

The JEL approach for the test is based on estimating $U$-type functionals with parameters being involved in those constraints, and it is quite different from the JEL method in \cite{Jing2009} which are based on estimating $U$-statistics. Furthermore, nuisance parameter is also involved and we need to deal with an extra variation introduced by the plug-in estimator $\tilde{\gamma}$ of the nuisance parameter $\gamma(Y, X)$. 
Throughout the proof, we let $\bi z=(x, y)^T$ and $\bi Z_i=(X_i, Y_i)^T, i=1,2,...,n$.

\begin{lemma}\label{test}
Suppose that $\E X^2_1<\infty$, $\E Y^2_1<\infty$ and $\sigma^2_g(\Delta, \gamma_2)>0$.
Then, for each $\Delta$, as $n \to \infty$, we have
$$P(\min_{1\le i \le n}\hat{V}_{i}(\Delta) <0< \max_{1\le i \le n}\hat{V}_{i}(\Delta))\rightarrow 1,$$
where $\hat{V}_{i}(\Delta)$ is given by (\ref{pseudo}).
\end{lemma}
\textbf{Proof.}
 Define $w(\bi z; \Delta+\tilde{\gamma}_2)=(\Delta+\tilde{\gamma}_2)\E h_1(z, \bi Z_1)-\E h_2(z, \bi Z_1)$ and 
 $\psi(\bi z_1, \bi z_2;\Delta+\tilde{\gamma}_2)=g_1(\bi z_1, \bi z_2; \Delta+\tilde{\gamma}_2)-w(\bi z_1;\Delta+\tilde{\gamma}_2)-w(\bi z_2;\Delta+\tilde{\gamma}_2).$
By the Hoeffding decomposition 
$$M_{n}(\Delta)=2n^{-1}\sum_{i=1}^{n}w(\bi Z_i;\Delta+\tilde{\gamma}_2)+  {n \choose 2}^{-1} \sum_{i<j}\psi(\bi Z_i, \bi Z_j;\Delta+\tilde{\gamma}_2)$$
and after some simple algebra,
\begin{align*}
&\hat{V}_{i}(\Delta)=2w(\bi Z_i;\Delta+\tilde{\gamma}_2)+\frac{2}{n-2}\sum_{k=1, k\neq i}\psi(\bi Z_i, \bi Z_k;\Delta+\tilde{\gamma}_2)-{n-1 \choose 2}^{-1} \sum_{i<j}\psi(\bi Z_i, \bi Z_j;\Delta+\tilde{\gamma}_2) \\
&=2w(\bi Z_i; \gamma_2+\Delta)+(\tilde{\gamma}_2-\gamma_2) h_{2}(\bi Z_i, \bi Z_j)+\frac{2}{n-2}\sum_{k=1, k\neq i}\psi(\bi Z_i, \bi Z_k;\tilde{\gamma}_2+\Delta)\\ \nonumber
&-{n-1 \choose 2}^{-1} \sum_{i<j}\psi(\bi Z_i, \bi Z_j; \tilde{\gamma}_2+\Delta) \\
&:=2w(\bi Z_i;\gamma_2+\Delta)+(\tilde{\gamma}_2-\gamma_2) h_{2}(\bi Z_i, \bi Z_j)+R_{ni}(\Delta+\tilde{\gamma}_2).
\end{align*}
We have
\begin{align*}
\E [(\tilde{\gamma}_2-\gamma_2) h_{2}(\bi Z_i, \bi Z_j)]^2\to 0 \;\; \text{as $n$ $\to$  $\infty$},
\end{align*}
since $(\tilde{\gamma}_2-\gamma_2)=O(n^{-1/2})$ with a similar result of (\ref{3.5}) and $\max_{1\le i \neq j\le n}|h_{2}(\bi Z_i, \bi Z_j)|=o(n^{1/2})$ by Lemma A.4 of \cite{Jing2009}.
Further, by Lemma \ref{lemma},
\begin{align*}
&\E[\psi (\bi Z_1,\bi Z_2;\Delta+\tilde{\gamma}_2)]^2
=\E[h(\bi Z_1,\bi Z_2;\Delta+\tilde{\gamma}_2)-w(\bi Z_1;\Delta+\tilde{\gamma}_2)-w(\bi Z_2;\Delta+\tilde{\gamma}_2)]^2<\infty.
\end{align*}
Then
\begin{align*}
\E [R^2_{ni}(\Delta+\tilde{\gamma}_2)] \leq C n^{-1}\E [\psi (\bi Z_1,\bi Z_2;\Delta+\tilde{\gamma}_2)]^2+C n^{-2} \E [\psi (\bi Z_1,\bi Z_2;\Delta+\tilde{\gamma}_2)]^2  \rightarrow 0 \;\;\;  \text{as $n \to \infty,$}
\end{align*}
where $C$ is some generic constant. Hence, $R_{ni}(\Delta+\tilde{\gamma}_2) \to 0$ and $\hat{V}_{i}(\Delta) \to 2 w(\bi Z_i;\Delta+\tilde{\gamma}_2)$ in probability. 
Thus, with the same argument as the proof of Lemma A.1 in \cite{Jing2009}, as $n\to \infty$, $$P(\min_{1\le i \le n}\hat{V}_{i}(\Delta) <0< \max_{1\le i \le n}\hat{V}_{i}(\Delta))\rightarrow 1,$$ for every $\Delta$. $\hfill{\Box}$

\begin{lemma}\label{A.2}
If $\E X^2_1<\infty$ and $\E Y^2_1<\infty$, we have 
\begin{align*}
\sqrt{n} M_n(\Delta)/(2 \sigma_g(\Delta, \gamma_2))\stackrel{d} \rightarrow N(0, 1) \;\;\;as \;\;\; n\rightarrow \infty,
\end{align*}
where $M_n(\Delta)$ is given by (\ref{mn}).
\end{lemma}
\textbf{Proof.}
Let
\begin{align*} 
M^0_{n}(\Delta)=\frac{2}{n(n-1)} \sum_{1\leq i<j \leq n}g_1(\bi Z_i, \bi Z_j; \Delta+\gamma_2).
\end{align*}
We have
\begin{align*}
&M_n(\Delta)=[M_n(\Delta)-M^0_{n}(\Delta)]+M^0_n(\Delta).
\end{align*}
Obviously, $M^0_n(\Delta)$ is a $U$-statistic for true values $\Delta$ and $\gamma_2$, and hence $\sqrt{n} M^0_n(\Delta)/(2 \sigma_g(\Delta, \gamma_2))\stackrel{d} \rightarrow N(0, 1) \;\;\;as \;\;\; n\rightarrow \infty$. 
Furthermore, $M_n(\Delta)-M^0_{n}(\Delta)=(\gamma_2-\tilde{\gamma}_2)\frac{2}{n(n-1)} \sum_{1\leq i<j \leq n}h_2(\bi Z_i, \bi Z_j)$ which is negligible since  
$(\tilde{\gamma}_2-\gamma_2)=O(n^{-1/2})$ by (\ref{eqn:rhohat}).$\hfill{\Box}$
\begin{lemma}\label{11}
Let $S_n(\Delta)=n^{-1}\sum_{i=1}^n[\hat{V}_{i}(\Delta)]^2$, if $\E X^2_1<\infty$ and $\E Y^2_1<\infty$. Then with probability one, we have $S_n(\Delta)=4\sigma^2_g(\Delta, \gamma_2)+o(1)$ for every $\Delta$, where $\hat{V}_{i}(\Delta)$ is given by (\ref{pseudo}).
\end{lemma}
\textbf{Proof.}  
Let 
$\hat{V}^0_{i}(\Delta)$ denote (\ref{pseudo}) when $\tilde{\gamma}_2$ is replaced by the true value $\gamma_2$.
Define
$S^0_n(\Delta)=n^{-1}\sum_{i=1}^n [\hat{V}^0_{i}(\Delta)]^2$.
Then
\begin{align*}
S_n(\Delta)=S^0_n(\Delta)+[S_n(\Delta)-S^0_n(\Delta)].
\end{align*}
From Lemma A.3 in \cite{Jing2009}, $S^0_n(\Delta)=4\sigma^2_g(\Delta, \gamma_2)+o(1)$. We need to show that $S_n(\Delta)-S^0_n(\Delta)$ is negligble.
In fact, we can rewrite $M_{n}(\Delta)$ in (\ref{mn}) as 
\begin{align*}
&M_{n}(\Delta)=(\Delta+\tilde{\gamma}_2)\frac{2}{n(n-1)} \sum_{1\leq i<j \leq n}h_2(\bi Z_i, \bi Z_j)-\frac{2}{n(n-1)} \sum_{1\leq i<j \leq n}h_1(\bi Z_i, \bi Z_j)\\
&:=(\Delta+\tilde{\gamma}_2) U^2_{n}-U^1_{n},
\end{align*} 
where $U^i_{n}, i=1,2$, are $U$-statistics with kernels being $h_i(\cdot), i=1, 2,$ respectively. Define 
\begin{align*}
&\hat{V}^1_{i}=nU^1_{n}-(n-1)U^{1(-i)}_{n-1} \;\text{and}\; \hat{V}^2_{i}=nU^2_{n}-(n-1)U^{2(-i)}_{n-1}.
\end{align*}
Then
\begin{align*}
\hat{V}_{i}(\Delta)=(\Delta+\tilde{\gamma}_2)\hat{V}^2_{i}-\hat{V}^1_{i} 
\end{align*}
and thus
\begin{align*}
&S_n(\Delta)-S^0_n(\Delta)=\frac{1}{n}\sum_{i=1}^n\{[\hat{V}_{i}(\Delta)]^2-[\hat{V}^0_{i}(\Delta)]^2\}\\ \nonumber 
&=(\tilde{\gamma}_2-\gamma_{2})(2 \Delta+\tilde{\gamma}_2+\gamma_{2})\frac{1}{n}\sum_{i=1}^n [\hat{V}^1_{i}]^2-2(\tilde{\gamma}_2-\gamma_{2})\frac{1}{n}\sum_{i=1}^n [\hat{V}^2_{i}\hat{V}_{i}^1]\\ \nonumber 
&\to 0, \;\text{as $n \to \infty$}.
\end{align*}
This is true because 
\begin{align*}
&\frac{1}{n}\sum_{i=1}^n [\hat{V}^1_{i}]^2=\frac{1}{n}\sum_{i=1}^n (\hat{V}^1_{i}-\E [U^1_{n}]+ \E [U^1_{n}])^2\\
&=\frac{1}{n}\sum_{i=1}^n (\hat{V}^1_{i}-\E [U^1_{n}])^2+2 \E [U^1_{n}] \frac{1}{n}\sum_{i=1}^n \hat{V}^1_{i} 
- (\E [U^1_{n}])^2\\
&=\frac{1}{n}\sum_{i=1}^n (\hat{V}^1_{i}-\E [U^1_{n}])^2+2 U^1_{n} 
\E [U^1_{n}]- (\E [U^1_{n}])^2,
\end{align*}
in which  $1/n \sum_{i=1}^n (\hat{V}^1_{i}-\E [U^1_{n}])^2=C+o(1)$ for some constant $C$ by Lemma A.3 of \cite{Jing2009},  and $2 U^1_{n} 
\E [U^1_{n}]- (\E [U^1_{n}])^2$ goes to a finite number. That is,
$1/n \sum_{i=1}^n [\hat{V}^1_{i}]^2= C+o(1)$ for some constant $C$. We have a similar result for $1/n \sum_{i=1}^n [\hat{V}^2_{i}]^2$ and hence $1/n \sum_{i=1}^n [\hat{V}^2_{i}\hat{V}_{i}^1]<\infty$. In addition, $\tilde{\gamma}_2-\gamma_{2}=O(n^{-1/2})$. Thus $S_n(\Delta)-S^0_n(\Delta) \to 0$ for every $\Delta$.

$\hfill{\Box}$

\begin{lemma}
Let $H_n(\Delta,\tilde{\gamma}_2 )=\max_{1\le i \le n}|g_1(\bi Z_i,\bi Z_j;\Delta+\tilde{\gamma}_2)|$, if $\E X^2_1<\infty$ and $\E Y^2_1<\infty$. Then with probability one, $H_n(\Delta,\tilde{\gamma}_2 )=o(n^{1/2})$.
\end{lemma}
\textbf{Proof.}
\begin{align*}
&H_n(\Delta,\tilde{\gamma}_2 )=\max_{1\le i \le n}|(\Delta+\tilde{\gamma}_2) h_2(\bi Z_i,\bi Z_j)-h_1(\bi Z_i,\bi Z_j)| \\
&\leq 3 \max_{1\le i \le n}| h_2(\bi Z_i,\bi Z_j)|+\max_{1\le i \le n}| h_1(\bi Z_i,\bi Z_j)|\\
&=o(n^{1/2})+o(n^{1/2})=o(n^{1/2})
\end{align*}
by applying Lemma A.4 of \cite{Jing2009} to the functions $h_2$ and $h_1$.
$\hfill{\Box}$

\textbf{Proof of Theorem \ref{thtest}} 
Lemma \ref{test} guarantees the existence and uniqueness of the solution for (\ref{eqn:lambda}).
Applying the above lemmas and with similar arguments of the  proof of Theorem 1 in \cite{Jing2009}, we have 
$|\lambda(\Delta)|=O_p(n^{-1/2})$ and $\lambda(\Delta)=S_n^{-1} (\Delta)\frac{1}{n}\hat{V}_i(\Delta)+\beta=S_n^{-1}(\Delta)M_n(\Delta)+\beta$ for every $\Delta,$
where $\beta=o_p(n^{-1/2})$.
Then
\begin{align}\label{chidis}
-2\log R(\Delta)=\frac{n M^2_n(\Delta)}{S_n(\Delta)}-nS_n(\Delta)\beta^2+2\sum_{i=1}^{n}\eta_i,
\end{align}
where $\sum_{i=1}^{n}\eta_i=o_p(1)$.
By Lemma \ref{A.2} and Lemma \ref{11}, $\frac{n M^2_n(\Delta)}{S_n(\Delta)}\stackrel{d} \rightarrow \chi^2_1$, and the second and the last terms of (\ref{chidis}) are negligible. 
Theorem \ref{thtest} is proved. $\hfill{\Box}$

\bigskip
\bigskip

\medskip
\noindent
\textbf{Proof of Theorem \ref{wilkrhodif}} The theorem is dealing with a vector $U$-statistic type functional with kernel size $(2, 2)$. 

(i) $\bi U^{0}_{n_1, n_2}(\delta_1, \delta_2)= \ $the original statistics functionals, based on all observations;

(ii) $\bi U^{-i,0}_{n_1-1, n_2}(\delta_1, \delta_2)=  $the  statistics functionals, after leaving $\bi Z^{(1)}_i$ out, for $i=1,...,n_1$;

(iii) $\bi U^{0, -j}_{n_1, n_2-1}(\delta_1, \delta_2)=  $the  statistics functionals, after leaving $\bi Z^{(2)}_j$ out, for $j=1,...,n_2$,

and 
$\bi V_{i,0}(\delta_1, \delta_2)= n_1\bi U^{0}_{n_1, n_2}(\delta_1, \delta_2)-(n_1-1)\bi U^{-i,0}_{n_1-1, n_2}(\delta_1, \delta_2), i=1,...,n_1,$ 

$\bi V_{0,j}(\delta_1, \delta_2)= n_2 \bi U^{0}_{n_1, n_2}(\delta_1, \delta_2)-(n_2-1)\bi U^{0, -j}_{n_1, n_2-1}(\delta_1, \delta_2), j=1,...,n_2.$

By some simple algebra, we have
\begin{align*}
&n^{-1}_1\sum_{i=1}^{n_1}\bi V_{i,0}(\delta_1, \delta_2)=\bi U_{n_1, n_2}(\delta_1, \delta_2),\;\;
n^{-1}_2\sum_{j=1}^{n_2}\bi V_{0, j}(\delta_1, \delta_2)=\bi U_{n_1, n_2}(\delta_1, \delta_2)
\end{align*}
and
\[
\hat{\bi V}_{i}(\delta_1, \delta_2)=\left\{ \begin{array}{ll} 
\frac{n-1}{n_1-1}\bi V_{i,0}(\delta_1, \delta_2)-\frac{n_2}{n_1-1}\bi U_{n_1, n_2}(\delta_1, \delta_2)&\;\;  \mbox{if}\;\; 1 \leq i \leq n_1; \\\\
\frac{n-1}{n_2-1}\bi V_{0,i}(\delta_1, \delta_2)-\frac{n_1}{n_2-1}\bi U_{n_1, n_2}(\delta_1, \delta_2) & \;\;  \mbox{if}\;\; n_1+1 \leq i \leq n. 
\end{array} \right. 
\]

With the same argument of the proof of Theorem 2.1 in \cite{Zhao2016}, $E \hat{\bi V}_{i}(\delta_1, \delta_2)=\bi 0.$
Then Theorem  \ref{wilkrhodif} is proved by Theorem 1 of \cite{Li2016}.

%\section*{References}
%
%\bibliography{mybibfile}

\begin{thebibliography}{10}

\bibitem{Chen2008}
Chen, J., Variyath, A. M. and Abraham, B.  (2008). Adjusted empirical likelihood and its properties. {\em J. Comput.  Graph.  Statist.} {\bf 17} (2), 426-443.  


\bibitem{Feng2012}
Feng, H.  and Peng, L.  (2012). Jackknife empirical likelihood tests for distribution functions. {\em J. Statist. Plann. Inference } {\bf 142}, 1571-1585.




\bibitem{Hjort09}
Hjort, N.L, McKeague, N.L. and Van Keilegom, I. (2009). Extending the scope of empirical likelihood. {\em Ann. Stat. } {\bf 37} (3), 1079-1111. 




\bibitem{Jing2009}
Jing, B., Yuan, J. and Zhou, W.  (2009). Jackknife empirical likelihood. {\em J. Amer. Statist. Assoc.} {\bf 104}, 1224-1232.

\bibitem{Lai1999}
Lai, C.  (1999). Robustness of the sample correlation-the bivariate lognormal case. {\em J. Appl. Math. Decis. Sci. } {\bf 3} (1), 7-19.

\bibitem{Li2011}
Li, M., Peng, L. and Qi, Y. (2011).  Reduce computation in profile empirical likelihood method. {\em Canad. J. Statist.}  {\bf 39} (2),   370-384.

\bibitem{Li2016}
Li, Z., Xu, J. and Zhou, W.  (2016). On nonsmooth estimating functions via  jackknife empirical likelihood. {\em Scand. J. Stat.} {\bf 43}, 49-69.

\bibitem{Lich2013}
Lichman, M. (2013). UCI Machine Learning Repository [http://archive.ics.uci.edu/ml]. Irvine, CA: University of California, School of Information and Computer Science.


\bibitem{Liu2015}
Liu, Z., Xia, X. and Zhou, W.  (2015). A test for equality of two distributions via jackknife empirical likelihood and characteristic functions. {\em Comput. Statist. Data Anal.} {\bf 92}, 97-114.




\bibitem{Lohweg2013}
Lohweg, V., Hoffmann, J. L., D$\ddot{o}$rksen, H., Hildebrand, R., Gillich, E., Hofmann, J. and Schaede, J. (2013). Banknote Authentication with Mobile Devices. {\em Proc. SPIE 8665, Media Watermarking, Security, and Forensics.} 

\bibitem{Ma2012}
Ma, C. and Wang, X.  (2012). Application of the Gini correlation coefficient to infer regulatory relationships in transcriptome analysis. {\em Plant Physiol.} {\bf 160}, 192-203.






\bibitem{Owen1988}
Owen, A. (1988).  Empirical likelihood ratio confidence intervals  for  single functional. {\em Biometrika  }  {\bf 75},   237-249.

\bibitem{Owen1990}
Owen, A.  (1990).  Empirical likelihood ratio confidence regions. {\em Ann. Statist.}  {\bf 18},   90-120.

\bibitem{Peng2012}
Peng, L. (2012).  Approximate jackknife empirical likelihood method for estimating equations. {\em Canad. J. Statist.}  {\bf 40} (1),  110-123.

\bibitem{Qin1994}
Qin, J. and Lawless, J.F. (1994).  Empirical likelihood and general estimating equations . {\em Ann. Statist. }  {\bf 22},  300-325.

\bibitem{Sang2016}
Sang, Y., Dang, X. and Sang, H.  (2016).  Symmetric Gini covariance and correlation. {\em Canad. J. Statist.}  {\bf 44} (3),  323-342.


\bibitem{Schechtman87}
Schechtman, E. and Yitzhaki, S. (1987). A measure of association based on Gini's mean difference. {\em Comm. Statist.   Theory  Methods}  {\bf 16} (1),  207-231.



\bibitem{Schechtman99}
Schechtman, E. and Yitzhaki, S. (1999). On the proper bounds of the Gini correlation.{\em Econom. Lett.} {\bf 63},  133-138.

\bibitem{Schechtman03}
Schechtman, E. and Yitzhaki, S. (2003). A Family of Correlation Coefficients Based on the Extended Gini Index.
{\em J. Econ. Inequal.} {\bf 1} (2),  129-146.

\bibitem{Schechtman07} 
Schechtman, E., Yizhaki, S. and Artsev, Y. (2007). The similarity between mean-variance and mean-Gini: Testing for equality of Gini correlations. {\em Advances in Investment Analysis and Portfolio Management (AIAPM)}, {\bf 3}, 103-128.


\bibitem{Shao96}
 Shao, J. and Tu, D. (1996). {\em The Jackknife and Bootstrap}. Springer, New York.

\bibitem{Zhao2016}
Wang, D. and Zhao, Y.  (2016).  Jackknife empirical likelihood for comparing two Gini indices. {\em Canad. J. Statist.}  {\bf 44} (1),  102-119.


\bibitem{Wang2013}
Wang, R., Peng, L. and Qi, Y.  (2013).  Jackknife empirical likelihood test for equality of two high dimensional means. {\em Statist. Sinica}  {\bf 23},  667-690.




\bibitem{Witting95}
Witting, H. and M\"{u}ller-Funk, U. (1995). {\it Mathematische Statistik  II}, B.G. Teubner, Stuttgart.

\bibitem{Wood1996}
Wood, A.T.A., Do, K.A., Broom, N.M. (1996). Sequential linearization of empirical likelihood constraints with application to $U$-statistics. {\em J. Comput.  Graph.  Statist.} {\bf 5}, 365-385.

\bibitem{Xu10}
Xu, W., Huang, Y.S.,  Niranjan, M.  and  Shen, M. (2010). Asymptotic mean and variance of Gini correlation for bivariate normal samples. {\em IEEE Trans.  Signal Process.} {\bf 58} (2),  522-534.

\bibitem{Yitzhaki13}
Yitzhaki, S. and Schechtman, E.(2013). {\em The Gini Methodology}. Springer, New York. 
 
\end{thebibliography}
%
%\begin{thebibliography}{99}

\end{document}